\documentclass[twocolumn]{aastex63}
\received{... 2020}
\submitjournal{ApJ}

\shorttitle{CMF of the Massive Clump G33.92+0.11}
\shortauthors{Su\'arez et al.}
\graphicspath{{./}{figures/}}

\usepackage[caption=false]{subfig} 
\usepackage[referable]{threeparttablex} 
\usepackage{booktabs} 
\usepackage{tabularx}
\usepackage{amsmath}
\usepackage{multirow} 
\makeatletter
\renewcommand\paragraph{\@startsection{paragraph}{4}{\z@}%
            {-2.5ex\@plus -1ex \@minus -.25ex}%
            {1.25ex \@plus .25ex}%
            {\centering\itshape\normalsize}}
\makeatother
\begin{document}

\title{A Core Mass Function Indistinguishable from the Salpeter Stellar\\ Initial Mass Function Using 1000 au Resolution ALMA Observations}

\correspondingauthor{Genaro Su\'arez}
\email{gsuarez@uwo.ca}

\author[0000-0002-2011-4924]{Genaro Su\'arez}
\affil{Department of Physics and Astronomy, The University of Western Ontario, 1151 Richmond St, London, ON N6A 3K7, Canada}
\affil{Instituto de Radioastronom\'ia y Astrof\'isica, Universidad Nacional Aut\'onoma de M\'exico, Morelia, Michoac\'an 58089, M\'exico}

\author[0000-0002-0786-7307]{Roberto Galv\'an-Madrid}
\affil{Instituto de Radioastronom\'ia y Astrof\'isica, Universidad Nacional Aut\'onoma de M\'exico, Morelia, Michoac\'an 58089, M\'exico}

\author{Luis Aguilar}
\affil{Instituto de Astronom\'ia, Universidad Nacional Aut\'onoma de M\'exico, Unidad Acad\'emica en Ensenada, Ensenada BC 22860, M\'exico}

\author[0000-0001-6431-9633]{Adam Ginsburg}
\affil{Department of Astronomy, University of Florida, P.O. Box 112055, USA}

\author[0000-0002-2996-305X]{Sundar Srinivasan}
\affil{Instituto de Radioastronom\'ia y Astrof\'isica, Universidad Nacional Aut\'onoma de M\'exico, Morelia, Michoac\'an 58089, M\'exico}

\author[0000-0003-2300-2626]{Hauyu Baobab Liu}
\affil{Academia Sinica Institute of Astronomy and Astrophysics, P.O. Box 23-141, Taipei 10617, Taiwan}

\author[0000-0001-8600-4798]{Carlos G. Rom\'an-Z\'u\~niga}
\affil{Instituto de Astronom\'ia, Universidad Nacional Aut\'onoma de M\'exico, Unidad Acad\'emica en Ensenada, Ensenada BC 22860, M\'exico}



\begin{abstract}
We present the core mass function (CMF) of the massive star-forming clump G33.92+0.11 using 1.3~mm observations obtained with the Atacama Large Millimeter/submillimeter Array (ALMA). With a resolution of 1000~au, this is one of the highest resolution CMF measurements to date. The CMF is corrected by flux and number incompleteness to obtain a sample that is complete for gas masses $M\gtrsim2.0\ M_\odot$. The resulting CMF is well represented by a power-law function ($dN/d\log M\propto M^\Gamma$), whose slope is determined using two different approaches: $i)$ by least-squares fitting of power-law functions to the flux- and number-corrected CMF, and $ii)$ by comparing the observed CMF to simulated samples with similar incompleteness. We provide a prescription to quantify and correct a flattening bias affecting the slope fits in the first approach, which is caused by small-sample or edge effects when the data is represented by either classical histograms or a kernel density estimate, respectively. The resulting slopes from both approaches are in good agreement each other, with $\Gamma=-1.11_{-0.11}^{+0.12}$ being our adopted value. Although this slope appears to be slightly flatter than the Salpeter slope $\Gamma=-1.35$ for the stellar initial mass function (IMF), we find from Monte Carlo simulations that the CMF in G33.92+0.11 is statistically indistinguishable from the Salpeter representation of the stellar IMF. Our results are consistent with the idea that the form of the IMF is inherited from the CMF, at least at high masses and when the latter is observed at high-enough resolution.
\end{abstract}
    


\section{Introduction}
\label{sec:introduction}
The origin of the stellar initial mass function (IMF) is one the most important open questions in Astrophysics. Modern numerical simulations and semi-analitical models seem to be able to produce a mass spectrum of collapsed objects within molecular clouds (MCs) that resembles the shape of the stellar IMF, especially its slope at high masses \citep[e.g.,][]{Padoan-Nordlund2002,Hennebelle-Chabrier2008,Bonnell_etal2011,Krumholz_etal2012,BallesterosParedes_etal2015}. However, it is not clear which physical mechanisms are the most relevant: fragmentation (turbulent, thermal, or hierarchical), the accretion history of the collapsed objects, or radiative and mechanical feedback \citep[for recent reviews, see][]{Vazquez-Semadeni_etal2019, Lee_etal2020}. The molecular ISM is organized in a hierarchy of structures. The smallest entities are the so-called ``cores'', with sizes $< 0.1$ pc and H$_2$ densities $> 10^5$ cm$^{-3}$, and believed to be the immediate precursors to single stars or small stellar systems \citep[see the reviews of ][for the cases of low- and high-mass star formation, respectively]{Bergin-Tafalla2007, Motte_etal2018b}. Determinations of their mass distribution (the core mass function, or CMF) could hold the key to the origin of the stellar IMF \citep[e.g.;][]{Motte_etal1998,Alves_etal2007,Konyves_etal2015,Cheng_etal2018,Massi_etal2019,Takemura_etal2021}.    

The shape of the IMF, and whether it varies with environment, has substantial implications for wide ranging fields of Astrophysics \citep[e.g.,][]{Offner_etal2014, Hopkins2018}. Current determinations of the CMF enable direct comparisons to the IMF across the high-mass regime  \citep[e.g.;][]{Motte_etal2018,Sadaghiani_etal2020,Oneill_etal2021}, where the stellar IMF is better constrained \citep[][]{Bastian_etal2010}. Accordingly, determinations of the CMF of diverse star-forming regions using high-quality data are required to study the connection between the IMF and the CMF. Moreover, high-mass cores reside in clustered environments. Recent studies have reported examples of massive cores that appeared to be single when observed at $\sim$0.05~pc resolution, which then fragment into several sources and streamers when observed in finer detail \citep[e.g.,][]{Izquierdo_etal2018,Palau_etal2018,Beuther_etal2018}. \citet{Sadaghiani_etal2020} pointed out the relevance of having high resolution ($\sim$0.01~pc) data for resolving cores in high-mass, clustered environments.

G33.92+0.11 (herafter G33.92) is a high-mass star formation region that was found to have a gas velocity dispersion much smaller than virial \citep{Watt-Mundy1999,Liu_etal2012}. The recent study of \citet{Wang_etal2020} showed that on scales of several pc the region is indeed dominated by gravity. Interferometric studies found that the central pc-scale clump is flattened, Toomre-unstable, and fragments into several tens of cores, all embedded within arm-like structures  \citep{Liu_etal2012,Liu_etal2015,Liu_etal2019}.

In this study we use high-resolution ($\approx$1000~au at 7.1~kpc) Atacama Large Millimeter/submillimeter Array (ALMA) observations of the central clump in G33.92 to derive its high-mass CMF. In Section \ref{sec:observations_method} we present the data, describe the method to identify cores, and explain how core masses are obtained. The G33.92 CMF and its best power-law representation are presented in Section \ref{sec:CMF}, as well as the comparisons to other CMFs and to the stellar IMF. We discuss our main results in Section \ref{sec:discussion} and summarize our work in Section \ref{sec:summary}.

\section{Data and Analysis Methods}
\label{sec:observations_method}

\subsection{Data}
\label{sec:data}
The data that we use were  originally presented in \citet{Liu_etal2015,Liu_etal2019}. We refer to those publications for more information on the  calibration and imaging. In particular, we analyze the 1.3-mm (224.5 GHz) dust continuum image presented in \citet{Liu_etal2019}. This image includes data from the ALMA 12-m array and the Atacama Compact Array (ACA), covering $(u,v)$ scales in the range $\sim 10$ to 2540 k$\lambda$. The synthesized beam has a FWHM size $\theta_{\rm{maj}}\times\theta_{\rm{min}}=0''.16\times0''.09$, corresponding to a physical resolution of $1136\times639$ au ($0.0055\times0.0031$ pc) at the distance of $7.1_{-1.3}^{+1.2}$ kpc \citep{Fish_etal2003}. 

G33.92 has a known ultracompact HII region that in principle could contribute free-free emission to the observed flux at 1.3 mm. \cite{Liu_etal2019} estimated such contribution from the simultaneously covered H$30\alpha$ line. We verified that the correction is effectively zero for all cores, except for one of our cores (ID 50; Table~\ref{tab:cores_params}), for which a 3\% flux correction was applied. Therefore, any systematic error in our CMF measurements from free-free emission is negligible.

\subsection{Identification of Cores}
\label{sec:dendrograms}
To identify substructures in the ALMA images we used the \textit{dendrogram} algorithm \citep{Rosolowsky_etal2008} as implemented in the \texttt{astrodendro} Python package\footnote{\url{http://www.dendrograms.org/}}. This algorithm tracks hierarchical structure over different scales and labels them as \textit{trunks} for the largest structures, \textit{branches} for structures internal to trunks that continue splitting into smaller structures, and \textit{leaves} for the structures at the smallest level of the hierarchy. The identification of such hierarchy has been pointed out as an important factor to take into account when analyzing a CMF \citep{Pineda_etal2009}. This hierarchical structure will also allow us to define appropriate regions to measure the background of cores. In addition, the dendrogram algorithm permits to define cores with irregular morphologies, which have been observed and could be due to further fragmentation or gas flows \citep[e.g.;][]{Izquierdo_etal2018}.

We performed the identification and cataloging in the image prior to primary-beam correction in order to have a close to uniform noise across the image. As some of the \texttt{astrodendro} parameters are commonly given in terms of the image noise, we first calculated the median absolute deviation (MAD; a measure of dispersion more robust to outliers than the standard deviation) of our observations by considering the regions of the image that do not belong to any (sub)structure identified in \citet{Liu_etal2019}. The resulting MAD is 20.5 $\mu$Jy beam$^{-1}$, which after applying the 1.483 factor for consistency with the standard deviation is somewhat larger (30.4 $\mu$Jy beam$^{-1}$) than the rms noise estimated by \citet[][25 $\mu$Jy beam$^{-1}$]{Liu_etal2019}. 

To run \texttt{astrodendro} we set the minimum intensity value that is considered to be significantly above the noise to $\texttt{min\_value} = 5\times\mathrm{MAD}$, the minimum significant increment between adjacent peaks to $\texttt{min\_delta} = 5\times\mathrm{MAD}$, and the minimum number of adjacent pixels that a peak needs to contain to the number of pixels in a synthesized beam, $\texttt{min\_npix} = 168$. These parameters are conservative enough to ensure that no spurious structures are added to our catalog, while allowing for the identification of all the clearly detectable sources.

\begin{figure}
	\centering
	\includegraphics[width=0.45\textwidth]{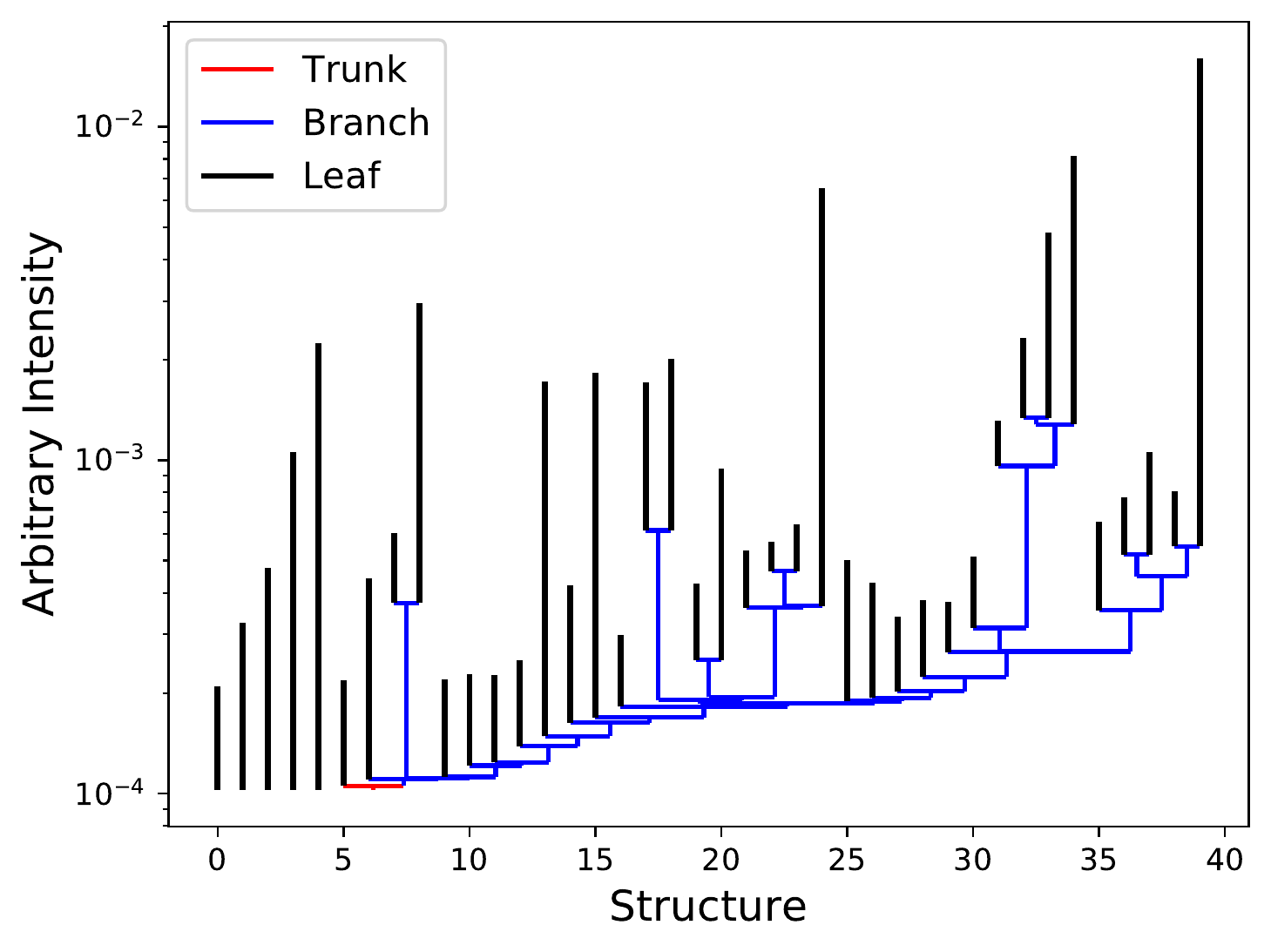}
	\caption{\textit{Dendrogram} hierarchy identified in the G33.92 clump. The trunk, branches, and leaves are plotted in red blue, and black, respectively. 74 (sub)structures were identified, of which 40 are leaves or cores.}
	\label{fig:dendrogram}
\end{figure}

\begin{figure*}
	\centering
	\includegraphics[width=1\textwidth]{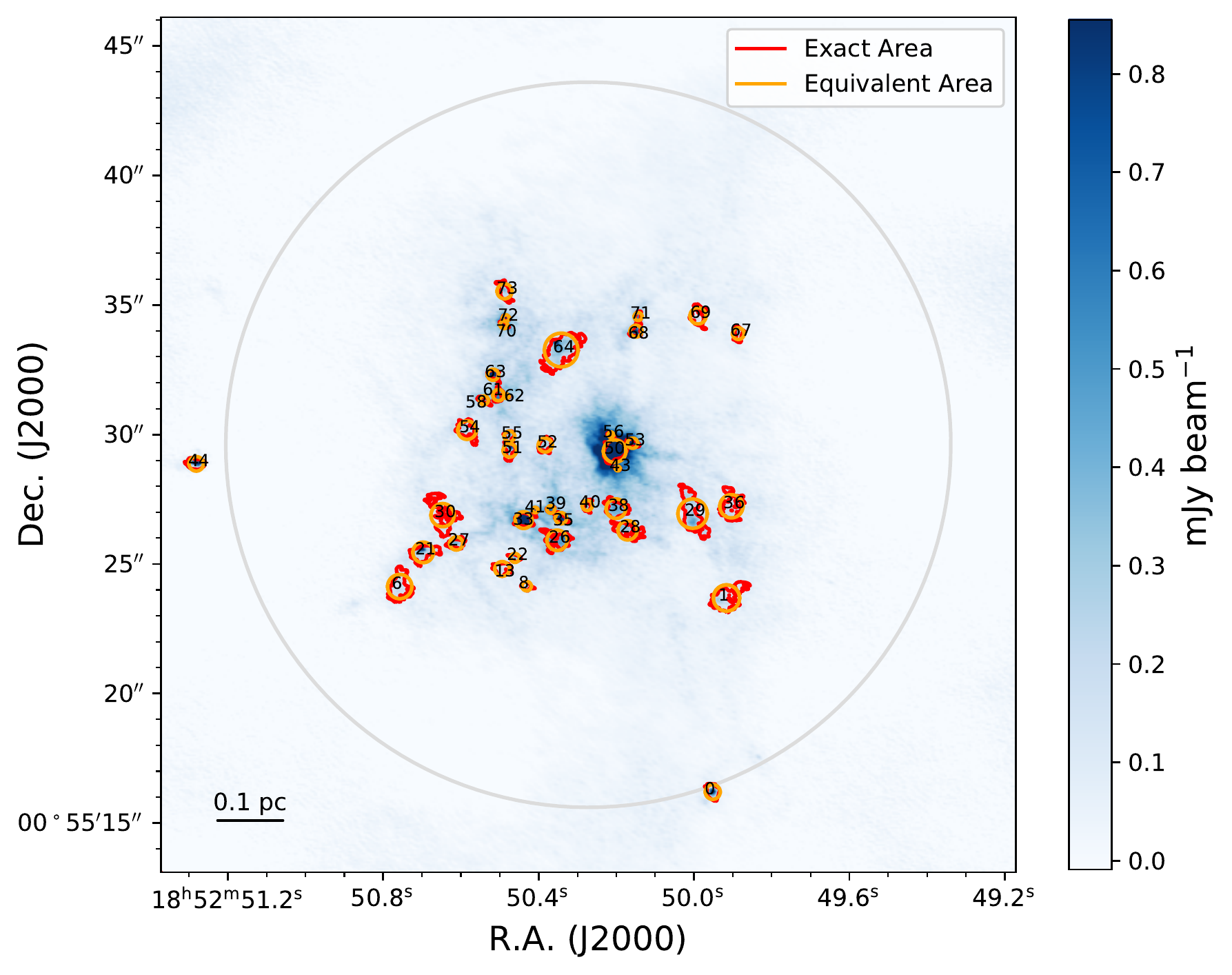}
	\caption{1.3 mm continuum image (blue map) overlaid with the identified cores (IDs from Table \ref{tab:cores_params}) in red contours. Their equivalent circular areas are shown as orange circles. The FWHM of the primary beam (28'') is indicated with the gray circle. The image is shown prior to primary-beam correction.}
	\label{fig:structure_cores}
\end{figure*}

In total 74 (sub)structures were identified, out of which 40 are leaves (the smallest element in the \textit{dendrogram} hierarchy), which we define as our {\it cores} \citep[e.g.;][]{Liu_etal2018,Lu_etal2020,Takemura_etal2021}. The hierarchy tree is shown in Figure~\ref{fig:dendrogram}. In Figure~\ref{fig:structure_cores} we show the ALMA dust-continuum image and the identified cores. We define the effective core radius as $r_c=\sqrt{A/\pi}$, where A is the exact area of the core provided by \texttt{astrodendro}. In Table \ref{tab:cores_params} we list the identified cores with their exact areas, ellipticities, and effective radii. Values for $r_c$ range between 600 and 4500 au, with a median of 1860~au. The median separation between cores is 0.22 pc, which is about 50 times the synthesized beam diameter. All cores but two have separations larger than $5$ synthesized beam diameters. Therefore, we deem our \textit{dendrogram}-based identification to be reliable.

Once the cores were identified, we calculated their fluxes to convert them into masses. We first computed the source-integrated fluxes in Jy. Then, we estimated their local background and subtracted it as follows: $i)$ for isolated cores (\textit{leaves} without a \textit{parent trunk}; 5 cores; Figure \ref{fig:dendrogram}) we defined the background as the average intensity in a surrounding ring with internal and external radius equal to $\times2$ and $\times3\ r_c$, respectively (we verified that elongated cores do not overlap with the defined rings); and $ii)$ for cores that are part of a hierarchical structure (35 cores; Figure \ref{fig:dendrogram}) we defined their background as the average intensity in their largest structures (\textit{ancestor trunks}) excluding all the smaller substructures (\textit{branches or leaves}) to avoid overestimation of their background. The core and background fluxes were obtained on the primary-beam corrected ALMA image. In Table \ref{tab:cores_params} we list the resulting fluxes as well as the flux uncertainties, which were defined as the square root of the number of beams in the core's area multiplied by the MAD noise level.

\subsection{Masses}
\label{sec:masses}
To derive the masses of the identified cores we use the optically-thin assumption. Only the central pixels of the few brightest sources have brightness temperatures $T_\mathrm{B}$ larger than a few~K, and only the central pixels of source ID 33 have $T_\mathrm{B} \approx 27$ K $\sim T_{\rm dust}$, which actually serves to validate our temperature estimation. The source-averaged brightness temperatures $T_\mathrm{B}$  before background subtraction  of the core sample range from 0.2 to 5.6~K. Therefore, the effects of potentially large optical depths are minimal, and we use the equation for optically-thin dust emission from a modified black body as: 

\begin{equation}
\label{eq:mass}
	M_{gas} = 26.6 \left(e^{\frac{11.1\textrm{ K}}{T_{\rm dust}}}-1 \right) \left(\frac{F_{1.3{\rm mm}}d^2}{\kappa_{1.3{\rm mm}}} \right),
\end{equation}

\noindent 
where the gas mass is in units of $M_\odot$ and has a $\times 100$ gas-to-dust conversion factor. $F_{1.3{\rm mm}}$ is the flux density at 1.3 mm in Jy,  $d=7.1_{-1.3}^{+1.2}$ kpc \citep{Fish_etal2003} is the distance to the object, and $\kappa_{1.3{\rm mm}}$ is the dust opacity, which we take as 0.6 cm$^2$g$^{-1}$ per unit of dust mass. The selected value for the dust opacity is typical for ISM dust in dense environments \citep{Ossenkopf-Henning_1994,Draine2006}, but (sub)mm opacities are uncertain within a factor of a few. Assuming a different fixed dust opacity would not change the shape of the resulting CMF,  but rather only produce a mass-scale shift. Opacity variations between cores could also occur due to core properties such as temperature, density, or evolutionary stage. We consider a 20\% error for the assumed opacity in our derived core masses. $T_{\rm dust}$ is the dust temperature, set equal to the dense-gas temperature profile of the clump given by \citet{Liu_etal2019}, with a small correction in the exponent and radical:

\begin{gather}
\label{eq:temperature}
	T(r) = 180\times \sqrt{0.00175/r} \times \omega(r) + 25 \times \left(1-\omega(r) \right)\\
	\omega(r)=e^{-r/0.0175} \nonumber
\end{gather}

\noindent 
Here $T(r)$ is in K units and $r$ is the radial distance in pc from the clump center, with coordinates R.A.$_{\rm J2000}=18^{\rm h}53^{\rm m}50^{\rm s}.204$, Dec$_{\rm J2000}=+00^\circ55'29''.3489$. This dense-gas temperature profile was determined from NH$_3$ $(J,K) = (1,1)$ to (3,3) hyperfine transitions for the outer part of the clump \citep{Liu_etal2012}, and from CH$_3$CN $J$=12--11 lines for the innermost area \citep{Liu_etal2015}. The temperature profile has an almost constant temperature of 25 K for $r\gtrsim0.05$ pc and a rapid increase for smaller $r$ values. In Table \ref{tab:cores_params} we list the radial distances of the identified cores and their corresponding temperatures with 10\% fractional uncertainties. Only the four central cores (IDs 43, 50, 53, and 56) have temperatures above 25~K (with a value of 318~K for the innermost core ID=50). The 36 remaining cores lie in the region of the clump where the temperature is almost constant.

We expect that our sample of cores is composed of prestellar and protostellar cores. The latter are expected to also have internal heating. However, with the available data it is difficult to separate both kind of cores to investigate how significant this effect could be. If the temperature of all cores is increased by 50\%, the resulting masses are smaller by a factor of 0.6. Nevertheless, high-resolution mid- and far-infrared data are needed to derive more accurate dust temperatures. Additionally, we expect that our cores are gravitationally bound. \citet{Liu_etal2015} showed (although at 0.5'' resolution) that all their detected cores in this region have velocity dispersions smaller than virial.

The derived gas masses of the cores are listed in Table~\ref{tab:cores_params} together with their uncertainties obtained by propagating the errors of the parameters in Equation \ref{eq:mass}, considering a 20\% uncertainty for the dust opacity. These masses range from 0.44 to 59~$M_\odot$. However, there are some issues that bias the detection of cores and their measured fluxes (and therefore their derived masses) as a function of their true masses. Hence, some corrections to remove these effects are needed before constructing a CMF.

\begin{table*}
\caption{Parameters of the Identified Cores.}
  \scriptsize
  \centering
  \label{tab:cores_params}
  \begin{threeparttable}
	\begin{tabularx}{\linewidth}{@{\extracolsep{+3pt}}cccccccccccc}
    \toprule
	ID  & R.A.$_{\rm J2000}$ & Dec.$_{\rm J2000}$                       & Area$^a$     & Ellipticity$^b$ & Radius$^c$ & Flux$_{\rm leaf}^d$ & Flux$_{\rm bg}^e$ & Flux$_{\rm core}^f$ & D$_{\rm cen}^g$ & $T$        & mass            \\
        & 18h52m(s)          & 00$^\circ$55$^\prime$($^{\prime\prime}$) & (arcsec$^2$) &                 & (pc)       & (mJy)               & (mJy)             & (mJy)               & (pc)        & (K)        & (M$_\odot$)     \\
    \midrule                                                                          
	0   & 49.95              & 16.2                                     & 0.257        & 0.272           & 0.010      & 10.58           & 0.86          &   9.72$\pm$0.08 & 0.47        &  25$\pm$2  & 12.14$\pm$5.00  \\
    1   & 49.92              & 23.7                                     & 0.803        & 0.493           & 0.017      &  8.86           & 3.22          &   5.64$\pm$0.14 & 0.24        &  25$\pm$2  &  7.05$\pm$2.91  \\
    6   & 50.76              & 24.1                                     & 0.688        & 0.408           & 0.016      &  8.47           & 5.03          &   3.44$\pm$0.13 & 0.34        &  25$\pm$2  &  4.30$\pm$1.78  \\
    8   & 50.43              & 24.1                                     & 0.101        & 0.412           & 0.006      &  1.09           & 0.74          &   0.35$\pm$0.05 & 0.21        &  25$\pm$2  &  0.44$\pm$0.19  \\
   13   & 50.49              & 24.8                                     & 0.252        & 0.317           & 0.010      &  2.42           & 1.84          &   0.58$\pm$0.08 & 0.22        &  25$\pm$2  &  0.72$\pm$0.31  \\
   21   & 50.70              & 25.4                                     & 0.494        & 0.444           & 0.014      & 10.24           & 3.61          &   6.63$\pm$0.11 & 0.29        &  25$\pm$2  &  8.28$\pm$3.41  \\
   22   & 50.46              & 25.2                                     & 0.094        & 0.461           & 0.006      &  1.18           & 0.68          &   0.49$\pm$0.05 & 0.19        &  25$\pm$2  &  0.62$\pm$0.26  \\
   26   & 50.35              & 25.9                                     & 0.482        & 0.310           & 0.013      & 13.14           & 3.52          &   9.61$\pm$0.11 & 0.14        &  25$\pm$2  & 12.01$\pm$4.95  \\
   27   & 50.61              & 25.8                                     & 0.202        & 0.495           & 0.009      &  3.66           & 1.47          &   2.18$\pm$0.07 & 0.24        &  25$\pm$2  &  2.73$\pm$1.13  \\
   28   & 50.17              & 26.3                                     & 0.429        & 0.600           & 0.013      &  7.36           & 3.14          &   4.22$\pm$0.10 & 0.11        &  25$\pm$2  &  5.28$\pm$2.18  \\
   29   & 50.00              & 26.9                                     & 1.003        & 0.676           & 0.019      & 17.06           & 7.33          &   9.73$\pm$0.16 & 0.13        &  25$\pm$2  & 12.16$\pm$5.01  \\
   30   & 50.65              & 26.9                                     & 0.628        & 0.520           & 0.015      &  6.25           & 4.59          &   1.66$\pm$0.13 & 0.24        &  25$\pm$2  &  2.07$\pm$0.87  \\
   33   & 50.44              & 26.7                                     & 0.328        & 0.312           & 0.011      & 49.60           & 2.40          &  47.20$\pm$0.09 & 0.15        &  25$\pm$2  & 58.97$\pm$24.28 \\
   35   & 50.34              & 26.8                                     & 0.147        & 0.539           & 0.007      &  6.24           & 1.08          &   5.16$\pm$0.06 & 0.11        &  25$\pm$2  &  6.45$\pm$2.66  \\
   36   & 49.90              & 27.2                                     & 0.630        & 0.411           & 0.015      &  7.46           & 4.60          &   2.86$\pm$0.13 & 0.17        &  25$\pm$2  &  3.57$\pm$1.48  \\
   38   & 50.20              & 27.1                                     & 0.445        & 0.273           & 0.013      & 10.32           & 3.25          &   7.07$\pm$0.11 & 0.08        &  25$\pm$3  &  8.82$\pm$3.64  \\
   39   & 50.37              & 27.1                                     & 0.068        & 0.176           & 0.005      &  2.57           & 0.50          &   2.07$\pm$0.04 & 0.12        &  25$\pm$2  &  2.59$\pm$1.07  \\
   40   & 50.27              & 27.3                                     & 0.119        & 0.377           & 0.007      &  2.18           & 0.87          &   1.31$\pm$0.05 & 0.08        &  25$\pm$3  &  1.64$\pm$0.68  \\
   41   & 50.41              & 27.1                                     & 0.033        & 0.110           & 0.004      &  1.34           & 0.24          &   1.10$\pm$0.03 & 0.13        &  25$\pm$2  &  1.37$\pm$0.57  \\
   43   & 50.20              & 28.7                                     & 0.025        & 0.336           & 0.003      &  1.65           & 0.18          &   1.47$\pm$0.03 & 0.02        &  31$\pm$3  &  1.38$\pm$0.57  \\
   44   & 51.28              & 28.9                                     & 0.245        & 0.511           & 0.010      & 14.99           & 0.00          &  14.99$\pm$0.08 & 0.56        &  25$\pm$2  & 18.73$\pm$7.71  \\
   50   & 50.20              & 29.4                                     & 0.596        & 0.220           & 0.015      & 77.09           & 4.36          &  72.74$\pm$0.12 & 0.00        & 318$\pm$32 &  5.78$\pm$2.34  \\
   51   & 50.47              & 29.4                                     & 0.210        & 0.447           & 0.009      &  4.48           & 1.53          &   2.94$\pm$0.07 & 0.14        &  25$\pm$2  &  3.68$\pm$1.52  \\
   52   & 50.38              & 29.6                                     & 0.225        & 0.287           & 0.009      &  3.39           & 1.65          &   1.74$\pm$0.08 & 0.09        &  25$\pm$2  &  2.18$\pm$0.90  \\
   53   & 50.16              & 29.7                                     & 0.122        & 0.571           & 0.007      & 24.81           & 0.89          &  23.92$\pm$0.06 & 0.03        &  30$\pm$3  & 23.86$\pm$9.80  \\
   54   & 50.58              & 30.2                                     & 0.400        & 0.398           & 0.012      &  5.64           & 2.92          &   2.72$\pm$0.10 & 0.20        &  25$\pm$2  &  3.40$\pm$1.41  \\
   55   & 50.48              & 29.9                                     & 0.129        & 0.203           & 0.007      &  2.51           & 0.94          &   1.57$\pm$0.06 & 0.14        &  25$\pm$2  &  1.96$\pm$0.81  \\
   56   & 50.21              & 30.0                                     & 0.034        & 0.565           & 0.004      &  3.44           & 0.25          &   3.19$\pm$0.03 & 0.02        &  32$\pm$3  &  2.97$\pm$1.22  \\
   58   & 50.54              & 31.3                                     & 0.089        & 0.655           & 0.006      &  2.44           & 0.65          &   1.78$\pm$0.05 & 0.19        &  25$\pm$2  &  2.23$\pm$0.92  \\
   61   & 50.50              & 31.5                                     & 0.146        & 0.478           & 0.007      &  4.91           & 1.06          &   3.85$\pm$0.06 & 0.17        &  25$\pm$2  &  4.81$\pm$1.98  \\
   62   & 50.48              & 31.5                                     & 0.024        & 0.408           & 0.003      &  0.75           & 0.17          &   0.58$\pm$0.02 & 0.16        &  25$\pm$2  &  0.72$\pm$0.30  \\
   63   & 50.52              & 32.3                                     & 0.144        & 0.295           & 0.007      & 12.07           & 1.05          &  11.02$\pm$0.06 & 0.19        &  25$\pm$2  & 13.77$\pm$5.67  \\
   64   & 50.34              & 33.3                                     & 1.323        & 0.457           & 0.022      & 23.22           & 9.66          &  13.55$\pm$0.18 & 0.15        &  25$\pm$2  & 16.93$\pm$6.98  \\
   67   & 49.89              & 33.9                                     & 0.156        & 0.505           & 0.008      &  1.91           & 0.55          &   1.36$\pm$0.06 & 0.23        &  25$\pm$2  &  1.69$\pm$0.70  \\
   68   & 50.15              & 34.0                                     & 0.161        & 0.339           & 0.008      &  7.76           & 1.18          &   6.58$\pm$0.06 & 0.16        &  25$\pm$2  &  8.22$\pm$3.39  \\
   69   & 49.99              & 34.6                                     & 0.304        & 0.399           & 0.011      &  2.70           & 1.28          &   1.42$\pm$0.09 & 0.21        &  25$\pm$2  &  1.77$\pm$0.74  \\
   70   & 50.49              & 34.2                                     & 0.064        & 0.473           & 0.005      &  3.87           & 0.46          &   3.41$\pm$0.04 & 0.22        &  25$\pm$2  &  4.26$\pm$1.75  \\
   71   & 50.14              & 34.6                                     & 0.074        & 0.514           & 0.005      &  2.22           & 0.54          &   1.68$\pm$0.04 & 0.18        &  25$\pm$2  &  2.10$\pm$0.87  \\
   72   & 50.48              & 34.5                                     & 0.050        & 0.338           & 0.004      &  3.45           & 0.36          &   3.08$\pm$0.04 & 0.23        &  25$\pm$2  &  3.85$\pm$1.59  \\
   73   & 50.49              & 35.5                                     & 0.261        & 0.605           & 0.010      &  4.10           & 1.91          &   2.19$\pm$0.08 & 0.26        &  25$\pm$2  &  2.74$\pm$1.13  \\
    \bottomrule
	\end{tabularx}
	\begin{tablenotes}[para,flushleft]
		$^a$Exact area provided by \texttt{astrodendro}. \\
		$^b$Ellipticity=$1 - \frac{b}{a}$, where $a$ and $b$ are the major and minor semiaxes, respectively, provided by \texttt{astrodendro}. \\
		$^c$Effective core radius. \\
		$^d$Fluxes without background subtraction. \\
		$^e$Fluxes of the defined background regions. \\
		$^f$Fluxes after background subtraction. \\
		$^g$Distance toward the clump center. \\
	\end{tablenotes}
	(This table is available in machine-readable form.)
 \end{threeparttable}
\end{table*}

\subsection{Completeness Corrections}
\label{sec:corrections}
To calculate the fraction of cores of a certain mass and the fraction of their fluxes that are missed by the \textit{dendrogram} algorithm, we performed a Monte Carlo simulation. We generated 10$^4$ artificial elliptical Gaussian cores with FWHMs along the major and minor axes equal to the major ($a$) and minor ($b$) semiaxes obtained by randomly choosing ellipticities ($1-\frac{b}{a}$) and effective radii within the intervals defined by the detected cores (ellipticities 0.11--0.68, sizes 0.003--0.022 pc; Table~\ref{tab:cores_params}) and padding on either side of the ranges by a 10\% of the extreme values. The artificial cores were assigned with masses randomly drawn in logarithmic scale within the mass range of the detected cores (0.44--59 $M_\odot$) and considering the same padding. The artificial cores were also assigned with random positions within 0.56~pc of the clump center, which corresponds to the furthest detected core and is close to the edge of the image (Figure \ref{fig:structure_cores}).

\begin{figure}
\centering
\includegraphics[width=0.47\textwidth]{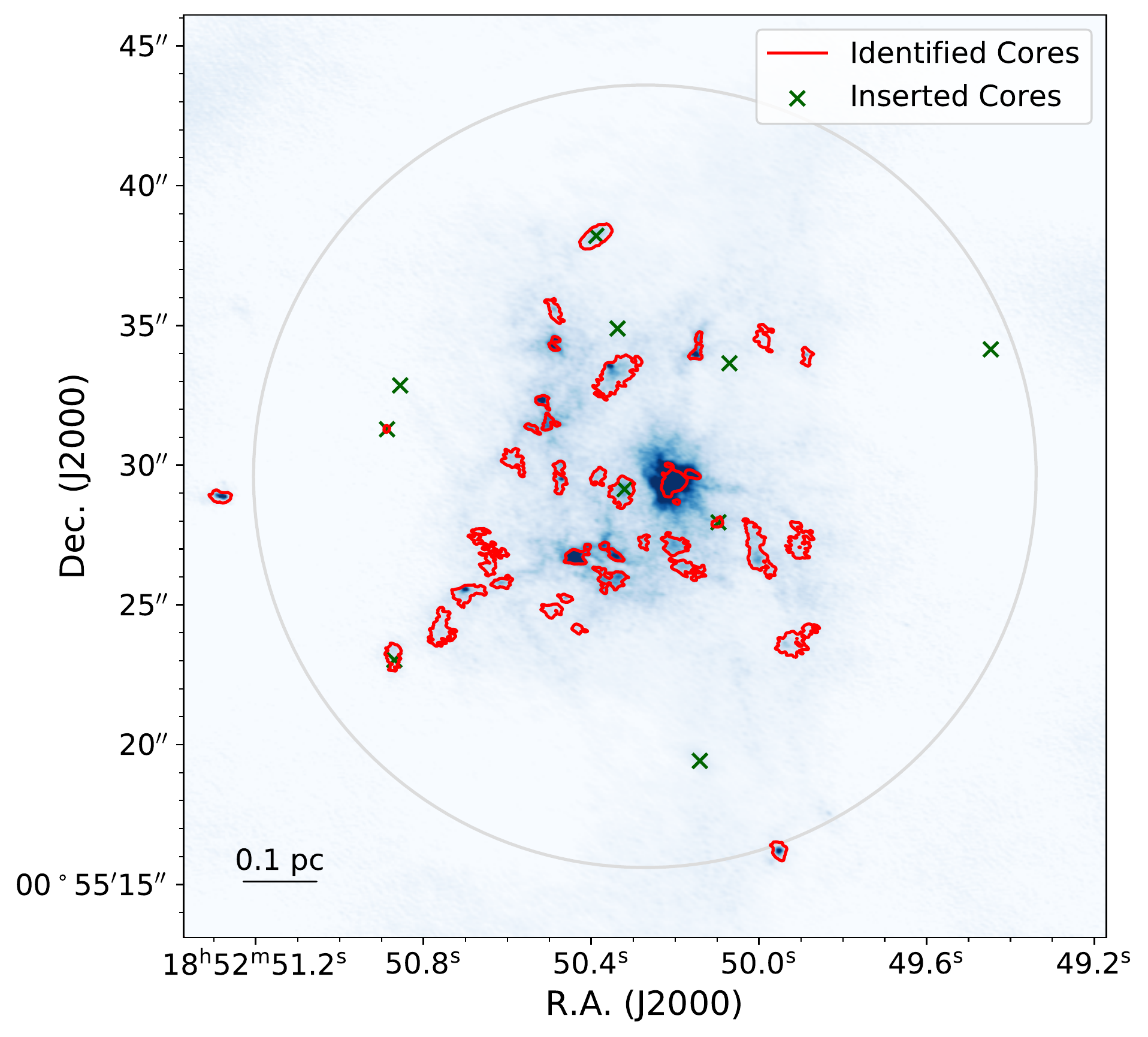}
\caption{Example of the insertion of elliptical Gaussian artificial cores (green crosses) into the ALMA image to evaluate the completeness of our sample of cores. The identified (real and artificial) cores are shown by the red contours. Five of the 10 artificial cores are recovered in this repetition. The color map and the gray circle are the same as in Figure \ref{fig:structure_cores}.}
\label{fig:structure_artificial}
\end{figure}

Before inserting the artificial cores into the ALMA image, the assigned masses were converted to fluxes by solving Equation \ref{eq:mass} and considering the temperature profile in Equation  \ref{eq:temperature}. These fluxes were then attenuated by the ALMA primary beam response. The insertion of the 10$^4$ artificial cores was done in groups of 10 at a time to avoid excessive overlaps. This leads to a correction for both sensitivity and confusion effects. An example of the insertion of artificial cores into the ALMA image is shown in Figure \ref{fig:structure_artificial}, where five out of 10 inserted cores were recovered by \texttt{astrodendro}.

\begin{figure}
	\centering
	\includegraphics[width=0.47\textwidth]{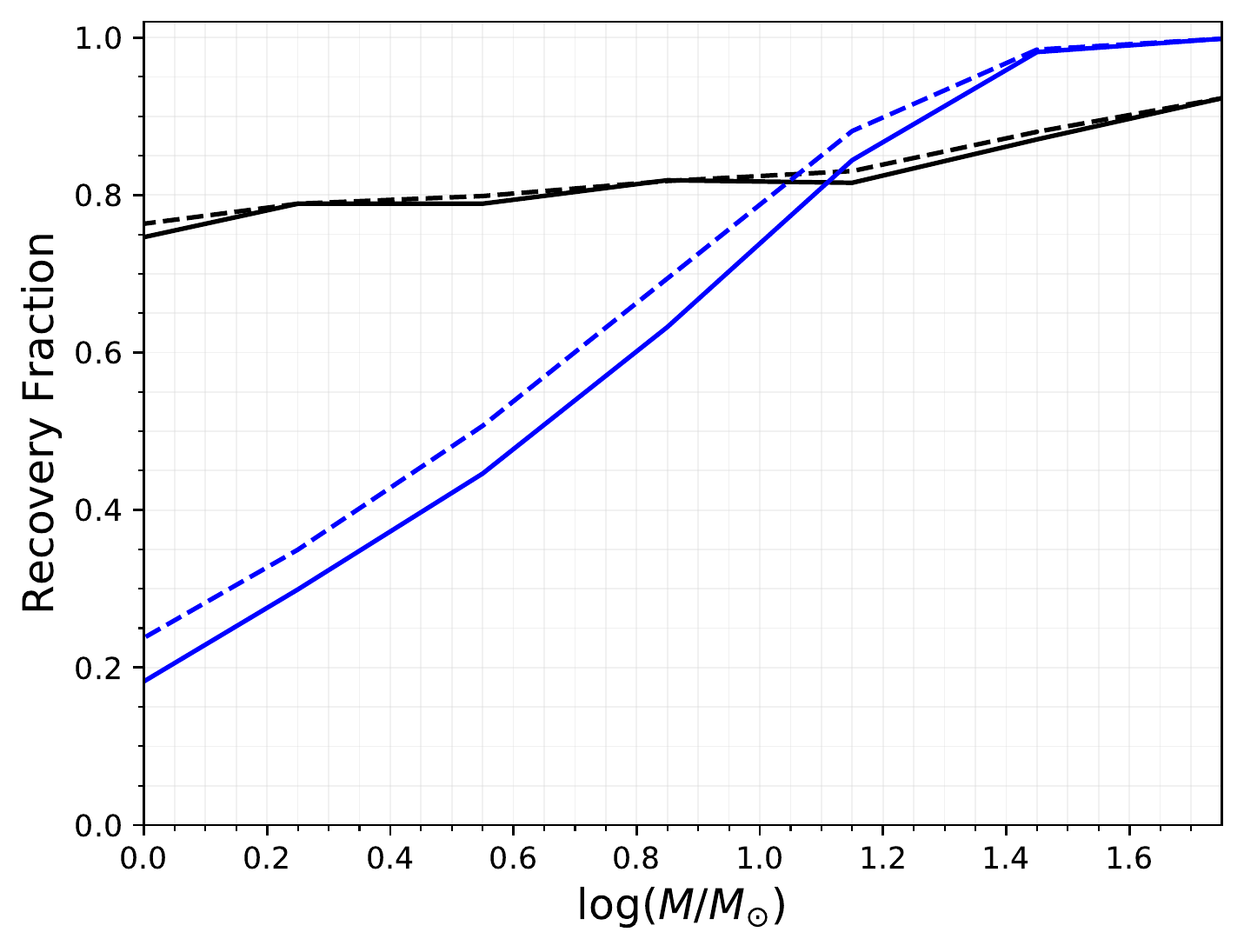}
	\caption{\textit{Flux recovery fraction} ($f_{\rm flux}$; black curves) and \textit{number recovery fraction} ($f_{\rm num}$; blue curves) as a function of both the input (solid curves) and the recovered (dashed curves) mass using our Monte Carlo simulation by the insertion and recovery of artificial cores.}
	\label{fig:CMF_corrections}
\end{figure}

We ran \texttt{astrodendro} with the same setup on the simulated images and applied the prescription explained in Section \ref{sec:dendrograms} to measure the background subtracted fluxes of the recovered cores. With this information we defined the median ratio of the recovered to input fluxes for cores of a given mass (or flux) as the \textit{flux recovery fraction} ($f_{\rm flux}$). The median ratio of the number of recovered cores to the number of inserted cores with a certain mass (or flux) was defined as the \textit{number recovery fraction} ($f_{\rm num}$). We derived these fractions as a function of both the input and the recovered mass. In Figure \ref{fig:CMF_corrections} we show the resulting $f_{\rm flux}$ and $f_{\rm num}$. For most of the mass range ($M\lesssim12\ M_\odot$) the effect of $f_{\rm flux}$ is smaller than that of $f_{\rm num}$, and the difference increases for lower masses, just for the massive cores the incompleteness due to $f_{\rm flux}$ are larger than those from $f_{\rm num}$. For example, the fraction of the flux of artificial cores with input masses $M$=5~$M_\odot$ ($\log (M/M_\odot)=0.7$) that is recovered by the \textit{dendrogram} algorithm is $\approx$80\%, while the probability of recovering cores with such masses is $\approx$55\%.

\section{Core Mass Function}
\label{sec:CMF}
The distribution of the identified cores masses in the ALMA observations corresponds to the observed or \textit{raw} CMF of G33.92. However, this distribution is affected by incompleteness due to the flux fraction and the number fraction of cores with a certain mass that are missed (Section \ref{sec:corrections}).

In this section we aim to obtain the high-mass slope of the \textit{corrected} CMF, i.e., the \textit{raw} CMF corrected by $f_{\rm flux}$ and $f_{\rm num}$ (Figure \ref{fig:CMF_corrections}). To achieve this purpose we follow two different approaches: $i)$ power-law fits to the \textit{corrected} CMF, whose resulting slopes were corrected by biases introduced due to a small sample size or edge effects, depending on the way the distribution is represented (Section \ref{sec:slope_fits}), and $ii)$ by comparing the \textit{raw} CMF to a number of simulated samples with incompleteness similar to our observations (Section \ref{sec:KS_test}).

\subsection{Slope Fits to the \textit{Corrected} CMF}
\label{sec:slope_fits}
\begin{figure*}
	\centering
	\subfloat[]{\includegraphics[width=.5\linewidth]{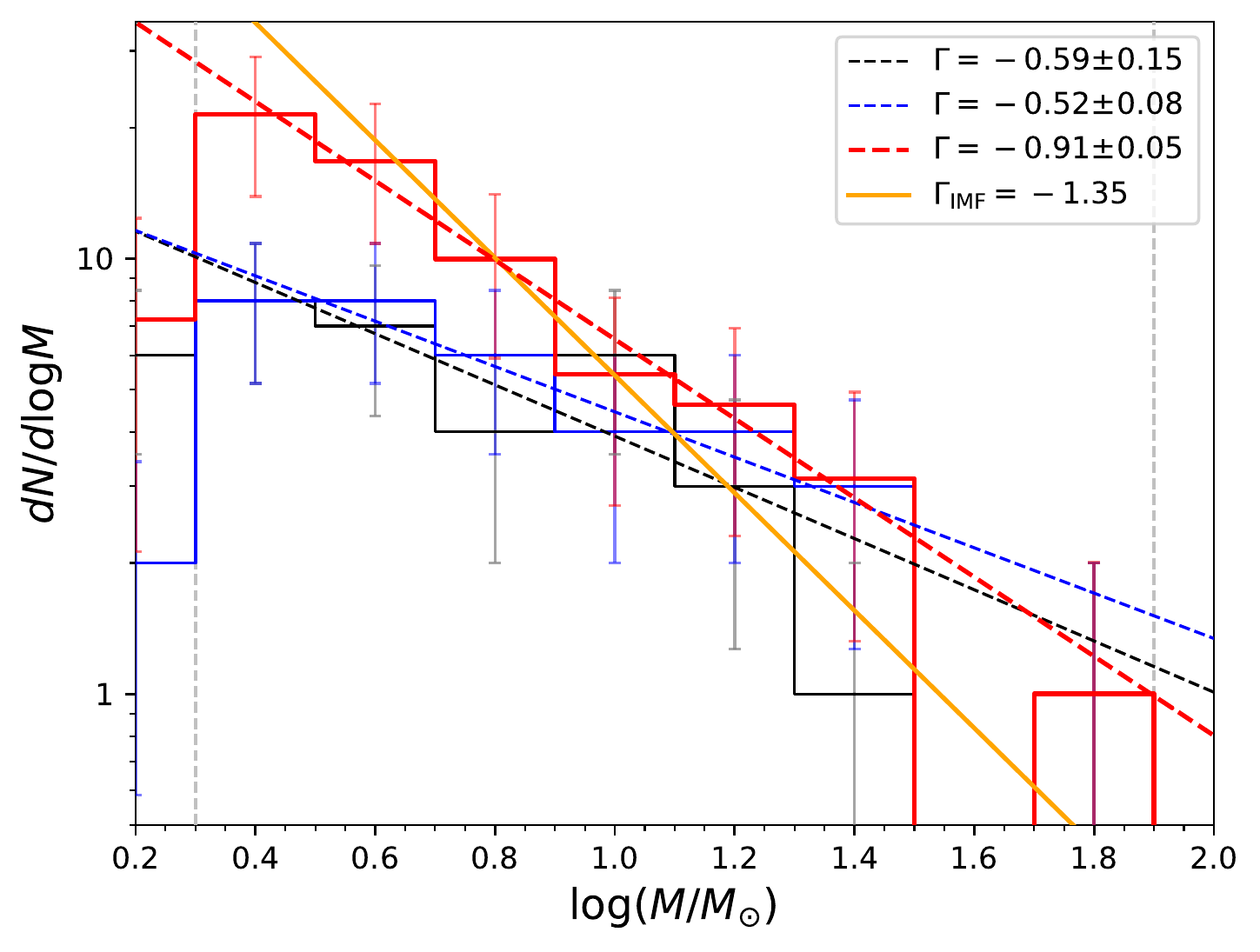}
	\label{fig:CMFa}}
	\subfloat[]{\includegraphics[width=.5\linewidth]{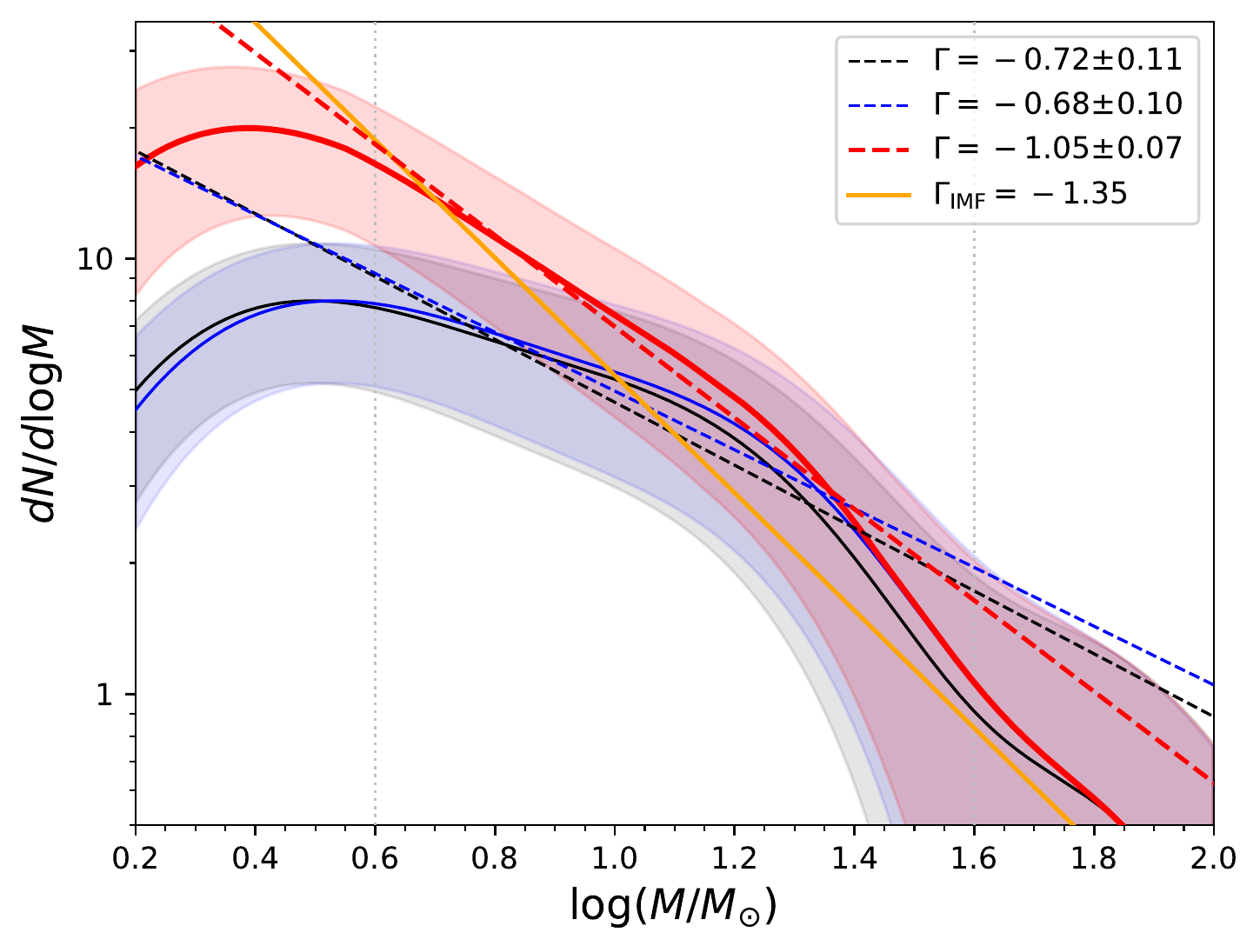}
	\label{fig:CMFb}}
	\caption{CMF of G33.92 based on classical histograms ({\bf a}) and KDEs ({\bf b}). Each panel shows the \textit{raw} CMF (black), the \textit{flux-corrected} CMF (blue), and the \textit{flux- and number-corrected} or just \textit{corrected} CMF (red). The best slope fits to the CMF as well as the high-mass IMF slope \citep[][]{Salpeter1955} are shown by the colored lines and have the values indicated in the legends. We further correct these slope measurements due to a flattening bias present in the fitting process (Sections \ref{sec:small-sample_correction} and \ref{sec:edge_effects}, and Figure \ref{fig:CMF_final}). The dashed gray lines (left panel) and the dotted gray lines (right panel) indicate the mass ranges considered for the slope fits to the histograms and KDEs, respectively. The uncertainties of the \textit{raw} CMF and the \textit{flux-corrected} CMF correspond to the Poisson errors ($\sqrt{N}$), and for the \textit{corrected} CMF we keep the fractional uncertainties of the \textit{flux-corrected} CMF. These errors are indicated by the vertical solid lines for the histograms and by the shadowed regions for the KDEs.}
	\label{fig:CMF}
\end{figure*}

We built the continuous \textit{raw} CMF using a kernel density estimation \citep[KDE, e.g.;][]{Silverman1986} considering a Gaussian kernel with a bandwidth of 0.2 dex, which corresponds to the average uncertainty of the core masses (Table \ref{tab:cores_params}). We also built the differential \textit{raw} CMF using classical histograms with a bin size of 0.2 dex. In Sections \ref{sec:small-sample_correction} and \ref{sec:edge_effects} we use Monte Carlo simulations to quantify and correct biases associated to the use of these representations. Similarly, we constructed the continuous and differential \textit{flux-corrected} CMF considering the core masses corrected by $f_{\rm flux}$ as a function of the recovered mass (Figure \ref{fig:CMF_corrections}). The \textit{flux-corrected} CMF was converted to the \textit{flux- and number-corrected} (or just \textit{corrected}) CMF by using $f_{\rm num}$ as a function of the input mass. In Figure \ref{fig:CMF} we show the CMF with these different levels of correction. The uncertainties of the \textit{raw} CMF and the \textit{flux-corrected} CMF are defined by the Poisson counting errors ($\sqrt{N}$), and for the \textit{corrected} CMF the fractional errors of the \textit{flux-corrected} CMF were kept.

We used an ordinary least-squares method to fit power-law distributions of the form $\frac{dN}{dlog(M)}\propto M^{\Gamma}$ to the CMF with different correction levels. The fits were done in the $\log N$--$\log M$ space and weighting by the inverse of the CMF uncertainties. The mass range considered for the slope fits to the histogram-based \textit{corrected} CMF is $0.3\le\log (M/M_\odot)\le1.9$, which corresponds to the range where this CMF has a clear power-law behaviour (Figure \ref{fig:CMF}). For the fits to the KDE form of the \textit{corrected} CMF we considered the range $0.6\le\log (M/M_\odot)\le1.6$, which is a good compromise between the amount of data used and the intensity of edge effects affecting the fits (see Section \ref{sec:edge_effects}). The slope values of the best power-law fits to the \textit{corrected} CMF are $\Gamma = -0.91\pm0.05$ and $\Gamma = -1.05\pm0.07$ based on histograms and KDEs, respectively. In Figure \ref{fig:CMF} we show the slope fits to the CMF before and after applying completeness corrections. As a reference, the \citet{Salpeter1955} stellar IMF slope for $M>1\ M_\odot$ is $-1.35$.

However, the slope fits to the \textit{corrected} CMF suffer from two different flattening effects depending on which representation is used; when working with histograms the effect is due to the small size of the sample, while for the case of KDEs it is caused by edge effects. We point out that although a maximum likelihood estimation (MLE) method may reduce these biases \citep{Clauset_etal2009,Koen-Kondlo2009}, it is not suitable for small-size samples like ours due to the restricted mass range where the data does not need completeness corrections. Consequently, the biases present in our least-squares fittings have to be quantified and corrected. In Sections \ref{sec:small-sample_correction} and \ref{sec:edge_effects} we used Monte Carlo simulations to address these issues.

\subsubsection{Small-Sample Correction for Histograms}
\label{sec:small-sample_correction}
In Section \ref{sec:slope_fits} we present the \textit{raw} CMF of G33.92 based on histograms as well as the slope fits to the \textit{corrected} CMF. However, a further correction must be done, in this case, to the value of the best slope fit to take into account an effect that flattens the slope. This bias arises because of small number statistics mainly affect the region with less counts, which in our case is the high mass range, making such slope fits a biased estimator of the real slope \citep[see also][]{MaizApellaniz-Ubeda2005}.

\begin{figure*}
	\subfloat[]{\includegraphics[width=0.5\linewidth]{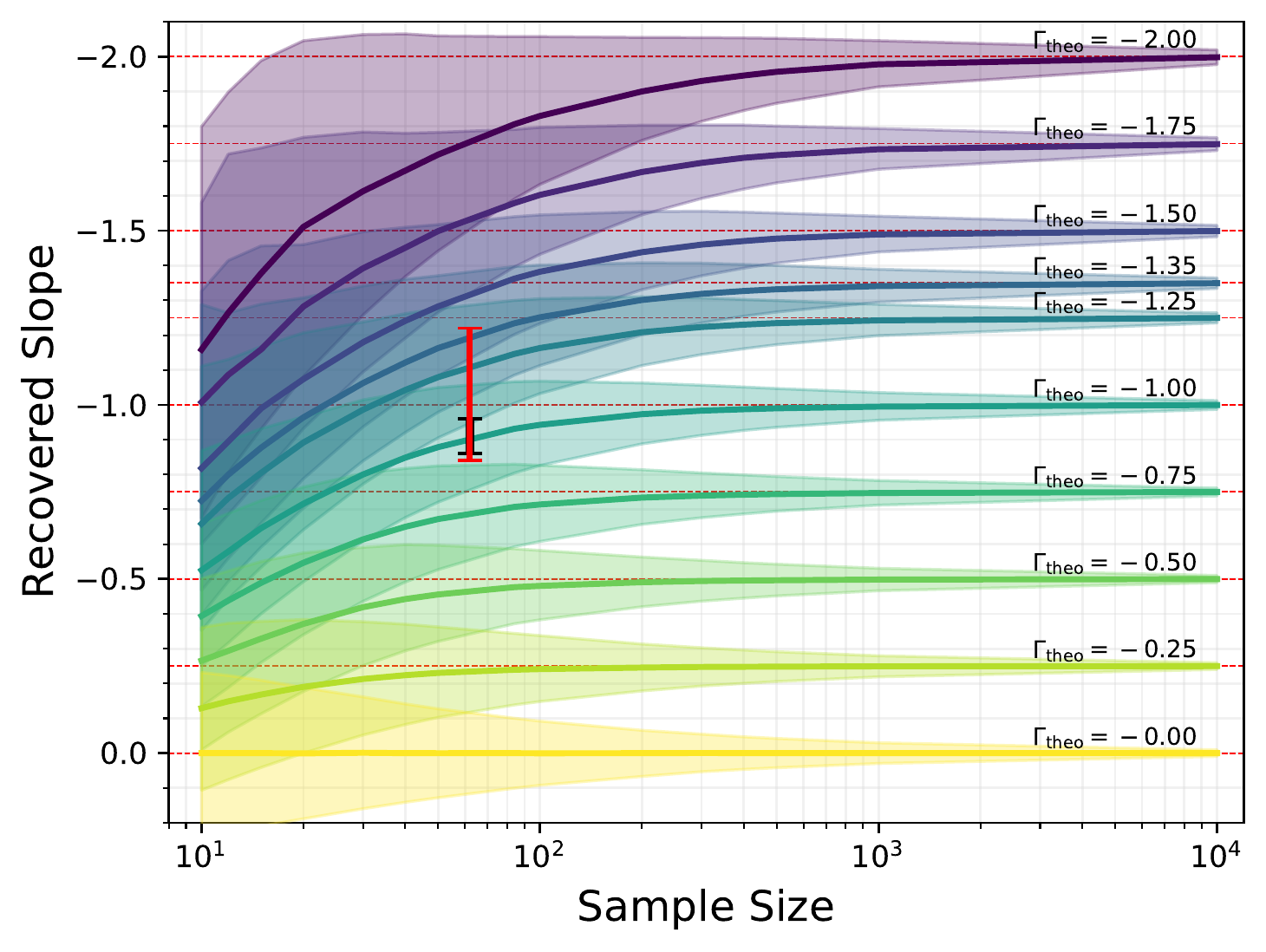} 
	\label{fig:slope_flattening_hist}}
	\subfloat[]{\includegraphics[width=0.5\linewidth]{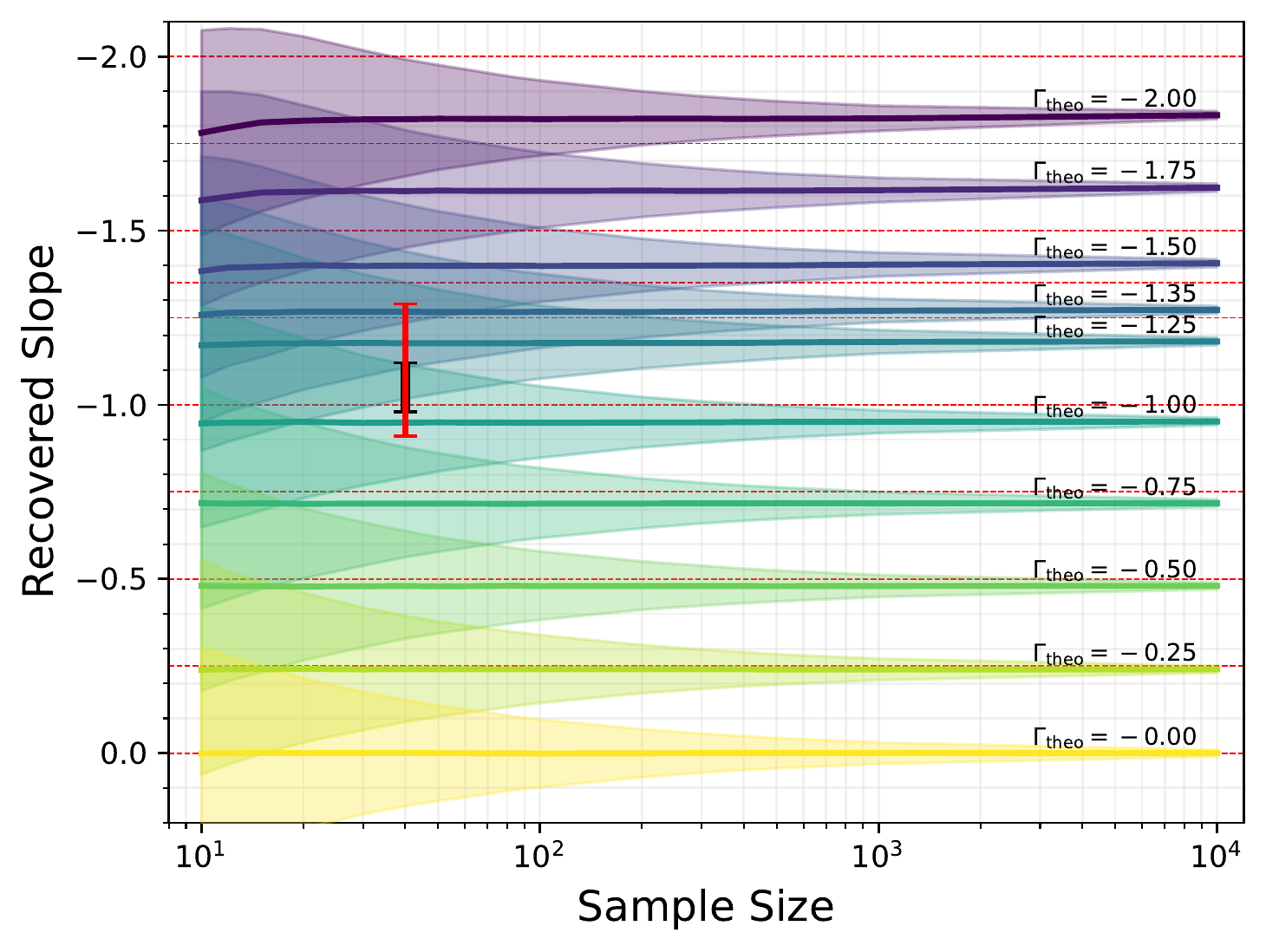}
	\label{fig:slope_flattening_KDE}}
	\caption{Recovered slopes from least-squares fits of power-law functions to histogram ({\bf a}) and KDE ({\bf b}) representations of mass distributions drawn from power-law distributions with the labelled theoretical slopes as a function of the sample size. The colored curves and shadowed regions represent, respectively, the median and the 68\% central confidence interval of the recovered slopes in the Monte Carlo simulation. As a reference, the horizontal red dashed lines indicate what the recovered slopes should be in the absence of the flattening bias. The black bar represents the best slope fit to the \textit{corrected} CMF before taking into account the flattening bias, and the red bar indicates the \textit{true} slope of the G33.92 CMF (after applying the correction due to this bias; Table \ref{tab:fitted_slopes}). The horizontal locations of these bars are determined by the total number of sources in the \textit{corrected} CMF in the mass ranges where the fits are done; 62 ($0.3\le\log (M/M_\odot)\le1.9$) and 40 ($0.6\le\log (M/M_\odot)\le1.6$) for the histogram and KDE forms, respectively.}
	\label{fig:slope_flattening}
\end{figure*}

To quantify the flattening bias we made random realizations taken from power-law distributions with theoretical slopes ($\Gamma_{\rm{theo}}$) ranging from $-$2 to 0 and with sample sizes ($N$) from 10 to 10$^4$ for each $\Gamma_{\rm{theo}}$. The masses were sampled in the same range used for the slope fits to the differential form of the \textit{corrected} CMF ($0.3\le\log (M/M_\odot)\le1.9$). We represented the distributions of each realization by histograms and fit power-law functions across the whole mass range in the same way as for the CMF (Section \ref{sec:slope_fits}). At each ($\Gamma_{\rm{theo}}$, $N$) combination we made 10$^5$ repetitions and consider the median and the 68\% central confidence interval, respectively, as the recovered slope and its uncertainty, as shown in Figure \ref{fig:slope_flattening_hist}.

We observe in Figure \ref{fig:slope_flattening_hist} that the recovered slopes flatten for smaller sample sizes (for $\Gamma_{\rm{theo}}\ne0$). This effect is stronger for steeper distributions. The origin of this effect is the lack of information in the high-mass regime as the sample size decreases. The fact that the effect diminishes as the theoretical slope approaches zero is because of the contrast between the amount of information in the low- and high-mass ranges also decreases. Thus, for a uniform distribution ($\Gamma_{\rm{theo}}=0$) this bias does not exist and the theoretical slope is always recovered regardless the sample size (only with changes in the slope errors).

\begin{figure*}
	\subfloat[]{\includegraphics[width=0.5\linewidth]{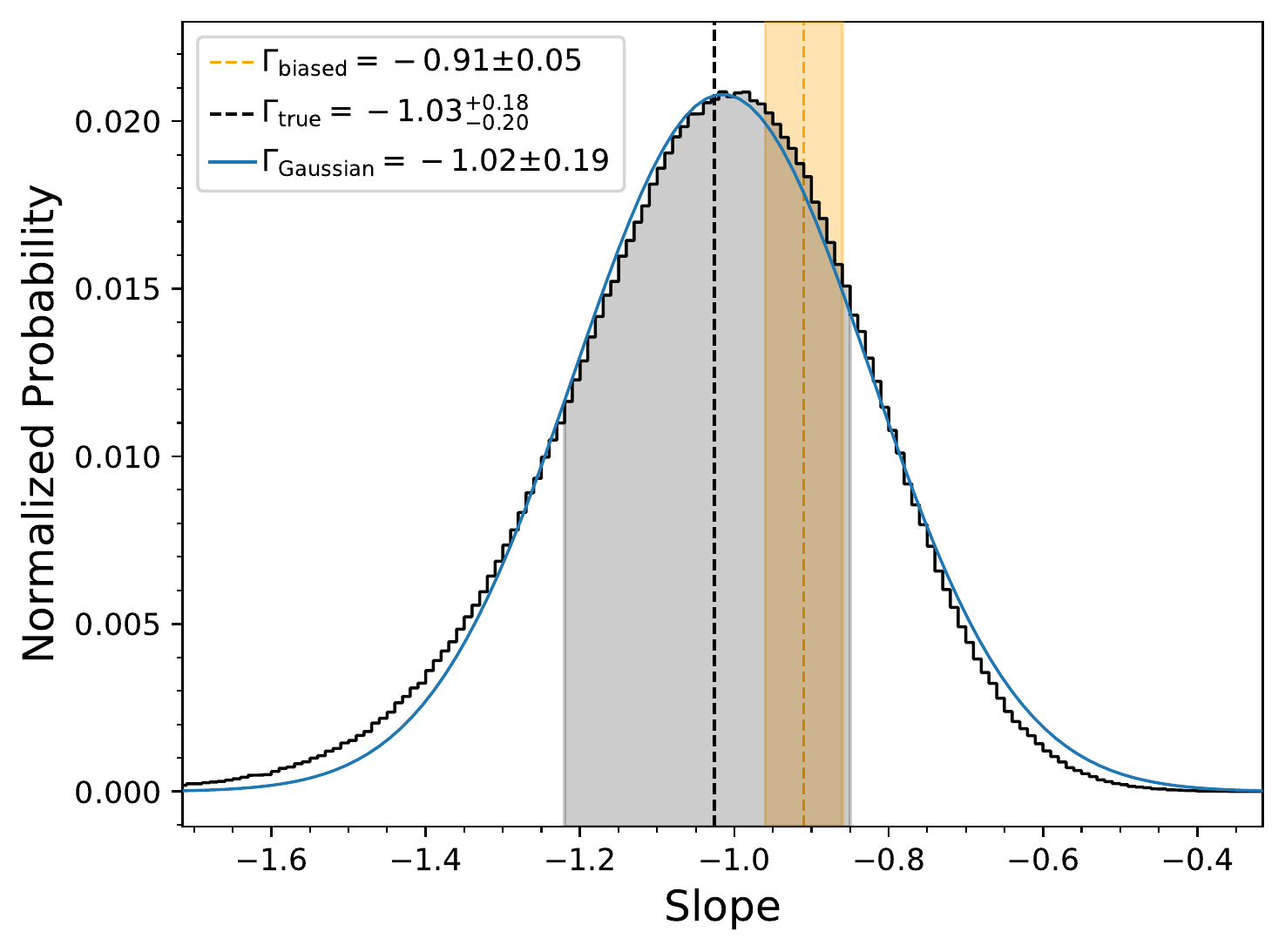} 
	\label{fig:PDF_histo}}
	\subfloat[]{\includegraphics[width=0.5\linewidth]{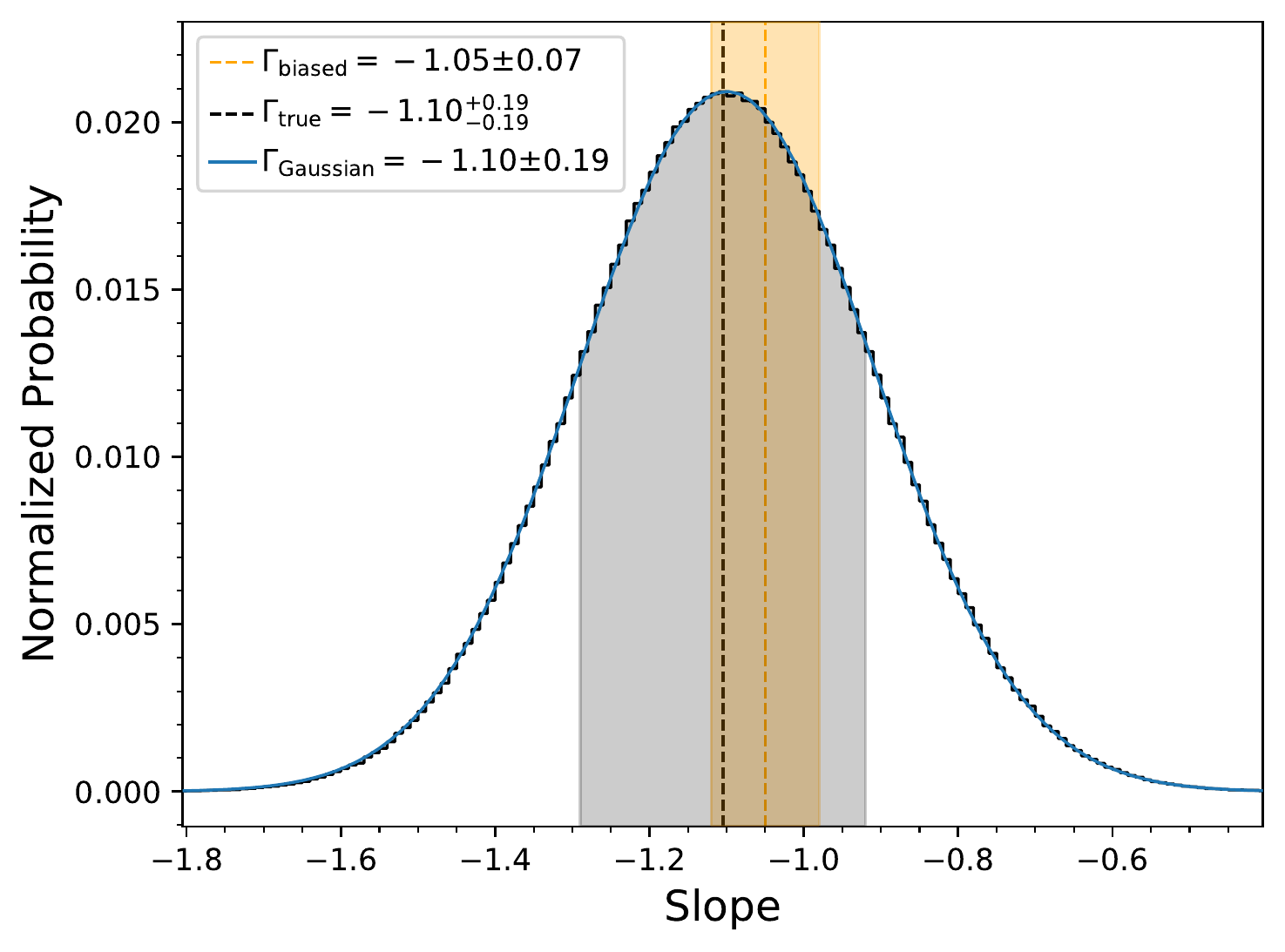}
	\label{fig:PDF_KDE}}
	\caption{Slope PDFs of the G33.92 CMF represented by histograms ({\bf a}) and KDEs ({\bf b}). The median and the 68\% central confidence interval of each PDF ($\Gamma_{{\rm true}}$) are indicated by the black dashed line and black shadowed region, respectively, and have the values indicated in the legend. Gaussian fits to the PDFs are represented by the blue curves and take the parameters (peak and standard deviation; $\Gamma_{{\rm Gaussian}}$) indicated in the legend. As a reference, the slope of the best power-law fit to the \textit{corrected} CMF ($\Gamma_{{\rm biased}}$) is shown by the orange dashed line and its uncertainty by the orange shadowed region.}
	\label{fig:PDF}
\end{figure*}

For a sample size of 62, which is the number of sources in the mass range ($0.3\le\log (M/M_\odot)\le1.9$) of the \textit{corrected} CMF where we fit a power-law function to its histogram representation, the flattening effect is important and must be taken into account. To statistically correct this bias we made random realizations as explained at the beginning of this section, but restricting $N$ to 62 sources and considering $\Gamma_{\rm{theo}}$ in the range from $-$2 to 0 in steps of 0.01. For a given $\Gamma_{\rm{theo}}$ we made 10$^5$ repetitions and recovered a slope in each by fitting a power-law function to the sampled distribution. To judge which $\Gamma_{\rm{theo}}$ produces the distribution with the recovered slope most consistent with the histogram-based \textit{corrected} CMF slope ($\Gamma = -0.91\pm0.05$; Table \ref{tab:fitted_slopes}), we first defined a Gaussian centered at this CMF slope and with a standard deviation equal to the slope error. We then defined the probability of a given $\Gamma_{\rm{theo}}$ for which the recovered slope is consistent with the slope of the histogram-based \textit{corrected} CMF as the value of the above mentioned Gaussian at the position of the recovered slope. The sum of the probabilities of the different $\Gamma_{\rm{theo}}$ was normalized to unity to obtain the probability distribution function (PDF) or inferred slope. In Figure \ref{fig:PDF_histo} we show the resulting PDF, which has a median and a 68\% central confidence interval of $\Gamma = -1.03_{-0.20}^{+0.18}$. We consider this inferred value as the \textit{true} slope of the G33.92 CMF based on histograms. In Table \ref{tab:fitted_slopes} we list the slope values from power-law fits to the different correction levels of the histogram-based CMF.

\subsubsection{Treatment of Edge Effects on KDEs}
\label{sec:edge_effects}
When the \textit{raw} CMF is represented by KDEs and then corrected by the flux and number incompleteness mentioned in Section \ref{sec:corrections}, power-law fits to the \textit{corrected} CMF also suffer of a systematic effect that flattens the slope. This bias is caused by edge effects on our KDE treatment due to our sample not being well characterized out of the mass range ($0.3\le\log (M/M_\odot)\le1.9$) used for the slope fits to the histogram-based \textit{corrected} CMF. Such issues characterizing the sample are due to its incompleteness for small masses ($\log (M/M_\odot)\lesssim0.2$) and the lack of information beyond the more massive detected core. 

We made random realizations in a fashion similar to that used to quantify the flattening effect when working with histograms (Section \ref{sec:small-sample_correction}). We represented the simulated distributions by KDEs and fit power-law functions in the same way as for the CMF (Section \ref{sec:slope_fits}). In Figure \ref{fig:slope_flattening_KDE} we show the results of this Monte Carlo simulation. We observe that the recovered slopes are systematically flatter than the theoretical ones (for $\Gamma_{\textrm{theo}}\ne0$). This effect is stronger for steeper distributions but, unlike the flattening effect when using histograms, it is almost insensitive to the size of the sample. Only for the steepest slopes ($\Gamma_{\textrm{theo}}<-1.5$) the bias slightly increases for sample sizes smaller than about 20 sources. 

The systematic flattening effect on slope fits to KDEs is minimized if the fits are done in a mass range considerably narrower than the range of the sampled masses. From our Monte Carlo simulation we found that for slope fits in the central half of the drawn masses' range the systematic effect flattens the recovered slope by about 2\% for $\Gamma_{\textrm{theo}}=-1.0$, while the effect could be up to 25\% if the fits are done across the whole mass range. In our case we generated masses in the range where the histogram-based \textit{corrected} CMF has a power-law behaviour ($0.3\le\log (M/M_\odot)\le1.9$) and made the slope fits in an interval diminished by 0.3 dex on both edges of the sampled masses, resulting in the range $0.6\le\log (M/M_\odot)\le1.6$. After trying several mass ranges in our simulations, we found this range is a good compromise between the magnitude of the effect and the amount of information used for the fits. We point out that, in order to recover the slope of a uniform distribution ($\Gamma_{\textrm{theo}}=0$), it is necessary to make the fits either in a symmetrically narrower region with respect to the interval where masses are drawn or across the entire mass range.

We made random realizations similar to those in Section~\ref{sec:small-sample_correction} to statistically correct the slope value of the best power-law fit to the KDE form of the \textit{corrected} CMF by the systematic flattening bias shown in Figure~\ref{fig:slope_flattening_KDE}. This time the random realizations had a sample size of 40, which is the number of sources in the mass range ($0.6\le\log (M/M_\odot)\le1.6$) of the \textit{corrected} CMF where we fit a power-law function to its KDE representation. The PDF was constructed as explained in Section \ref{sec:small-sample_correction}, with the Gaussian centered at the slope of the KDE-based \textit{corrected} CMF ($\Gamma = -1.05\pm0.07$; Table \ref{tab:fitted_slopes}) and with a standard deviation equal to the slope error. The median and the 68\% central confidence interval of the PDF are $-1.10_{-0.19}^{+0.19}$, as shown in Figure \ref{fig:PDF_KDE} and listed in Table \ref{tab:fitted_slopes}. We consider this inferred slope as the \textit{true} slope of the G33.92 CMF based on KDEs.

\subsection{Comparison of the \textit{raw} CMF to Simulated Incomplete Samples}
\label{sec:KS_test}
As an alternative method to that described in Section \ref{sec:slope_fits} for determining the \textit{true} slope of the G33.92 CMF, we present in this section an approach that avoids fitting power-law functions to the data. The idea of this approach is to simulate mass distributions with similar incompleteness to those affecting the \textit{raw} CMF and then look for the distribution that most closely resembles the data. 

We made random realizations from power-law distributions with theoretical slopes ranging from $-$2 to 0 in steps of 0.01 and with sample sizes of 30, which is the number of detected cores with masses $0.3\le\log (M/M_\odot)\le1.9$. In contrast with all the previous random realizations made in this study, for these particular ones we filtered each mass as it was drawn. First, the drawn mass was multiplied by the corresponding flux (or mass) completeness fraction according to $f_{\rm flux}$ (as a function of the input mass; Figure \ref{fig:CMF_corrections}). We then computed the number completeness fraction corresponding to the resulting mass considering $f_{\rm num}$ (as a function of the recovered mass; Figure \ref{fig:CMF_corrections}) and compared it to a generated random number between 0 and 1. If the random number is smaller than the number completeness fraction, then the mass is kept. Otherwise, the mass is rejected. We repeated the process until getting 30 masses for each realization. Thus, the resulting simulated distributions have similar incompleteness as the \textit{raw} CMF.

\begin{figure}
\centering
\includegraphics[width=0.45\textwidth]{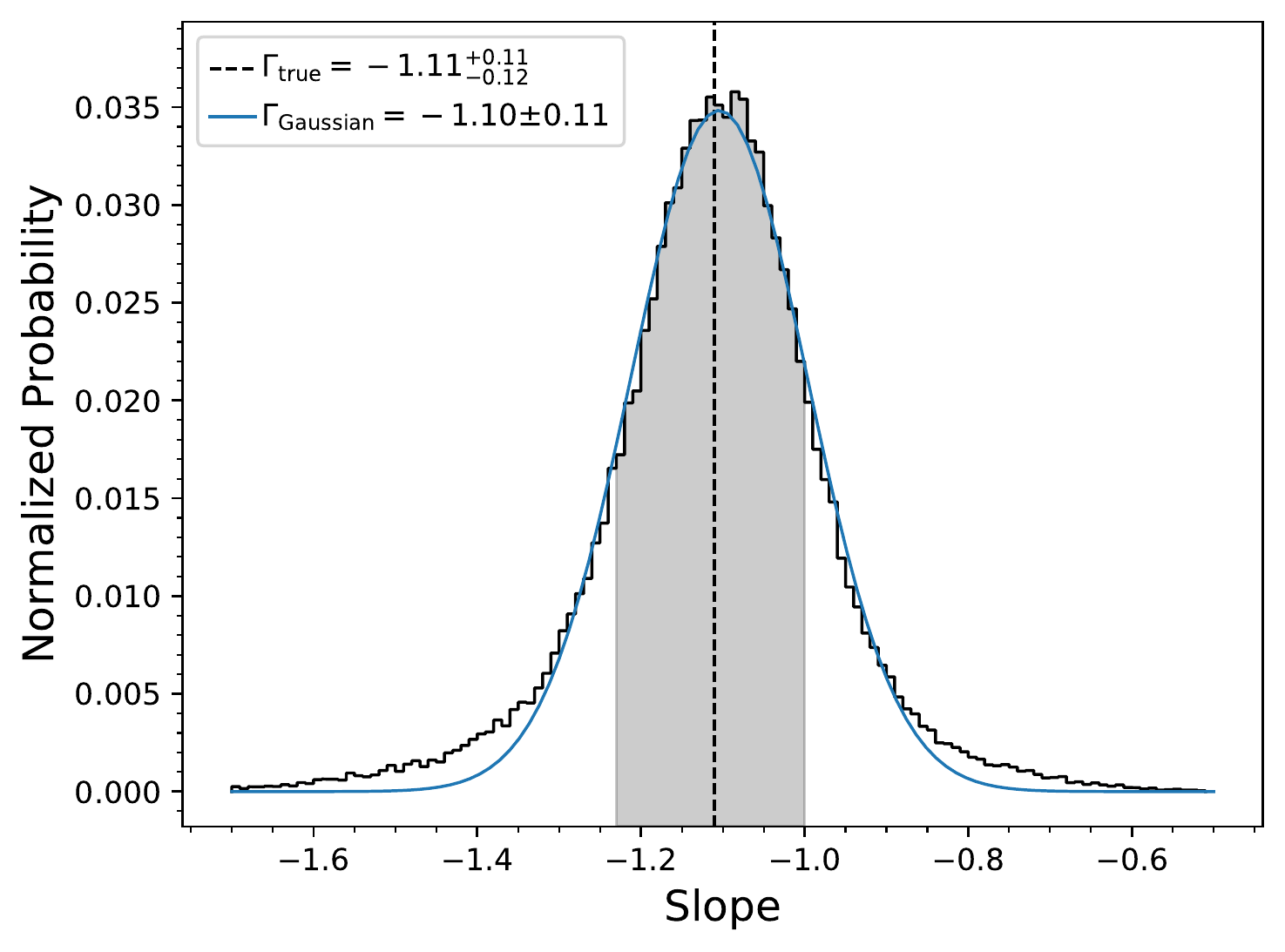}
\caption{Slope PDF of the G33.92 CMF based on KS tests between the \textit{raw} CMF and simulated incomplete samples. The median and the 68\% central confidence interval of the PDF ($\Gamma_{{\rm true}}$) are shown by the dashed line and shadowed region, respectively, and have the values indicated in the legend. The blue curve represents the best Gaussian fit to the PDF, which has the peak and the standard deviation ($\Gamma_{{\rm Gaussian}}$) indicated in the legend.}
\label{fig:PDF_KStest}
\end{figure}

The simulated incomplete mass distribution in every realization with a different theoretical slope was compared to the \textit{raw} CMF through a two-sample Kolmogorov-Smirnov (KS) test \citep{Press_etal1992}. This test assess the likelihood that two distributions are drawn from the same parent sample (null hypothesis). From all realizations, we selected the theoretical slope of the realization with the smallest KS statistic or the closest to unity $p$-value. If two or more realizations have the same KS statistic we choose the median slope. We repeated this process $10^5$ times to obtain the PDF of the theoretical slopes that generate the incomplete samples more similar to the \textit{raw} CMF. In Figure \ref{fig:PDF_KStest} we show the PDF or inferred slope, which has a median and a 68\% central confidence interval of -1.11$_{-0.12}^{+0.11}$. We consider this value as the \textit{true} slope of the G33.92 CMF from the comparison of the \textit{raw} CMF to simulated incomplete samples. In Table \ref{tab:fitted_slopes} we summarized the slope values derived in this study for the G33.92 CMF.

\begin{table}
\caption{Slope Representation of the G33.92 CMF.}
  \scriptsize
  \centering
  \label{tab:fitted_slopes}
  \begin{threeparttable}
	\begin{tabularx}{\linewidth}{@{\extracolsep{0pt}}lccc@{}}
    \toprule
	Correction Level                        & $\Gamma_{\rm {hist}}^a$ & $\Gamma_{\rm {KDE}}^b$   & $\Gamma_{\textrm {KS}}^c$ \\
    \midrule                                                                               
    \textit{Raw}$^d$                        & -0.59$\pm$0.15           & -0.72$\pm$0.11           & \nodata                  \\
    \textit{Flux-Corrected}                 & -0.52$\pm$0.08           & -0.69$\pm$0.10           & \nodata                  \\
    \textit{Flux- and Number-Corrected}$^e$ & -0.91$\pm$0.05           & -1.05$\pm$0.07           & \nodata                  \\
    \midrule                                                                      
    Inferred or \textit{True}$^f$           & -1.03$_{-0.20}^{+0.18}$  & -1.10$_{-0.19}^{+0.19}$  & -1.11$_{-0.12}^{+0.11}$  \\
    \bottomrule
	\end{tabularx}
	\begin{tablenotes}[para,flushleft]
		$^a$ Slope from power-law fits to the histogram-based CMF. \\
		$^b$ Slope from power-law fits to the KDE-based CMF. \\
		$^c$ Inferred slope from KS tests between the \textit{raw} CMF and simulated samples with incompleteness similar to the data. \\
		$^d$ Without applying corrections due to incompletenesses or biases introduced during the fitting process.\\
		$^e$ Also referred just as \textit{corrected} throughout the paper. \\
		$^f$ Median and 68\% central confidence interval of the inferred slope. For the cases when fits are done, this slope is after correcting by $f_{\rm flux}$ and $f_{\rm num}$, as well as by the flattening effects in the slope fits.
	\end{tablenotes}
 \end{threeparttable}
\end{table}

\begin{figure*}
	\centering
	\includegraphics[width=0.9\linewidth]{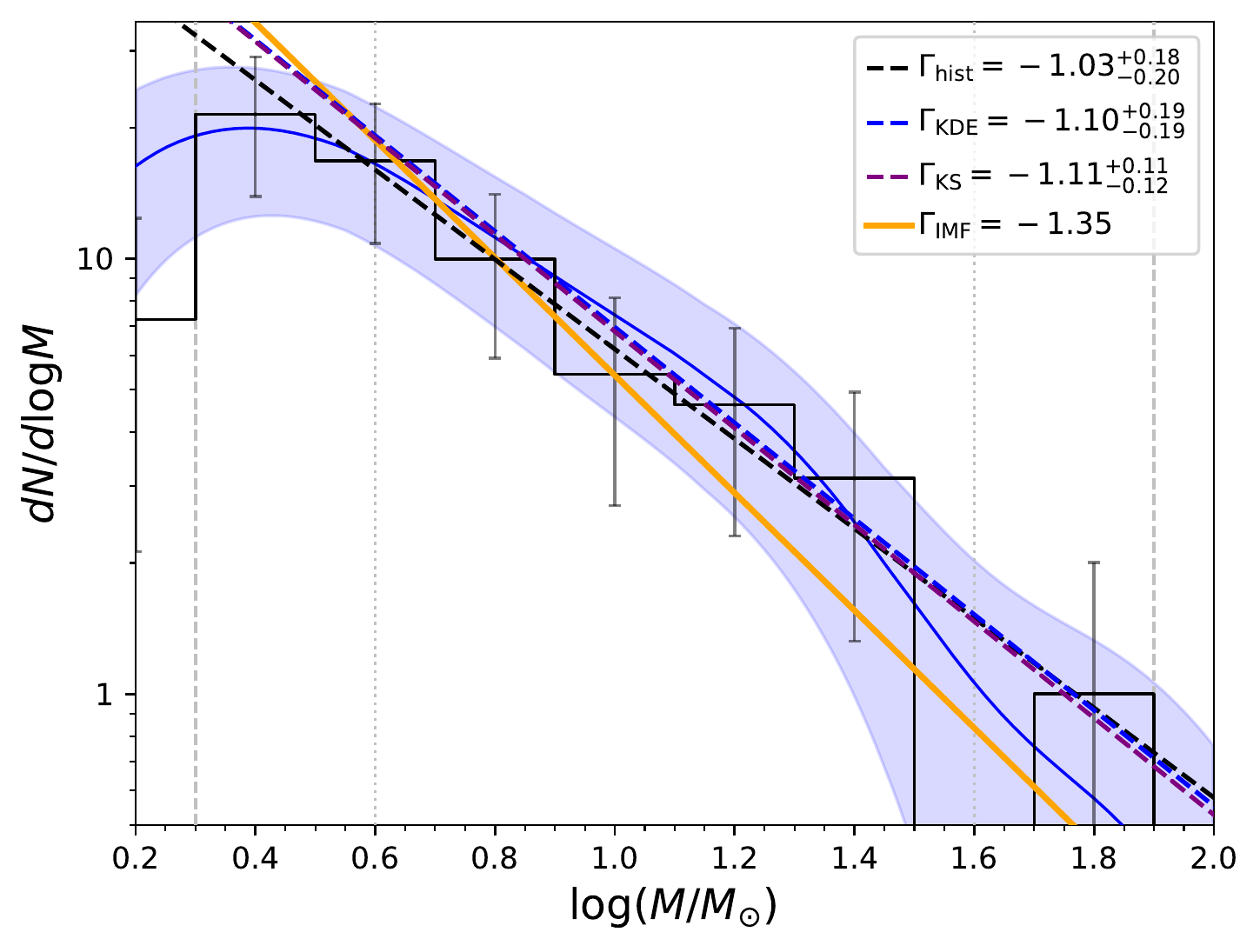}
	\caption{\textit{Corrected} CMF of G33.92 based on histograms (black) and KDEs (blue). The inferred or \textit{true} slopes derived in this study (Table \ref{tab:fitted_slopes}) are indicated by the dashed lines; the black and the blue lines from the slope fits to the histogram ($\Gamma_{\textrm{hist}}$) and KDE ($\Gamma_{\textrm{KDE}}$) representations of the  \textit{corrected} CMF, respectively, after correcting the flattening biases discussed in Sections \ref{sec:small-sample_correction} and \ref{sec:edge_effects}, and the purple line from comparisons, through a KS test, between the \textit{raw} CMF and simulated incomplete samples ($\Gamma_{\textrm{KS}}$; Section \ref{sec:KS_test}). The uncertainties and other lines are the same as in Figure \ref{fig:CMF}.}
	\label{fig:CMF_final}
\end{figure*}

\subsection{Comparisons to other CMFs and to the IMF}
\label{sec:comparisons_CMF_IMF}
The slopes of the high-mass CMF of G33.92 derived in this study following different approaches (Section \ref{sec:CMF}) through the use of differential distributions, continuous distributions, and KS test comparisons are in agreement with each other, as summarized in Table \ref{tab:fitted_slopes}. In Figure \ref{fig:CMF_final} we show the histogram- and KDE-based \textit{corrected} CMF together with the inferred or \textit{true} slope from each approach. The inferred slope from the comparison of the \textit{raw} CMF to simulated incomplete samples is about 60\% more precise than the inferred slopes from power-law fits to the \textit{corrected} CMF. Additionally, in the former case it is not necessary to use KDEs or histograms to represent the CMF since the comparisons are done through KS tests on the cumulative distributions. Thus, we adopt the slope from this approach, $\Gamma=-1.11_{-0.12}^{+0.11}$, as the slope of the G33.92 CMF.

The CMF of G33.92 is in general agreement with other high-mass CMFs derived from ALMA observations, e.g., W43-MM1 \citep[$\Gamma=-0.96\pm0.02$;][]{Motte_etal2018}, G286.21+0.17 \citep[$\Gamma=-1.24\pm0.17$;][]{Cheng_etal2018}, the IR-dark clumps of the ASHES survey \citep[$\Gamma=-1.07\pm0.09$;][]{Sanhueza_etal2019}, NGC6334 \citep[$\Gamma=-1.10\pm0.02$;][]{Sadaghiani_etal2020}, three clouds in the Central Molecular Zone of the Galaxy \citep[$\Gamma=-1.04\pm0.08$;][]{Lu_etal2020}, {and 28 dense clumps across the Galaxy \citep[$\Gamma=-0.94\pm0.08$;][]{Oneill_etal2021}}, but is somewhat steeper than the CMFs of another sample of IR-dark clumps  \citep[$\Gamma=-0.70\pm0.13$;][]{Liu_etal2018}, G28.37+0.07 \citep[$\Gamma=-0.87\pm0.07$;][]{Kong2019}, and G305.137+0.069 \citep[$\Gamma=0.02\pm0.37$;][]{Servajean_etal2019}, as listed in Table \ref{tab:cmf_comparisons}. Some of the CMFs in Table \ref{tab:cmf_comparisons} suggest an excess of high-mass cores with respect to the Salpeter IMF \citep[$\Gamma=-1.35$;][]{Salpeter1955}. All these CMFs, including ours, are expected to be a mixture of prestellar and protostellar cores, however, it is difficult to separate them because of the lack of mid- and far-infrared data at the angular resolution of the (sub)mm detections. \citet{Sanhueza_etal2019} and \citet{Nony_etal2020} used CO outflow activity to distinguish these two core populations, but our frequency setup does not cover the 2--1 line.

At face value, the final G33.92 CMF slope appears to be slightly flatter than the Salpeter IMF by about $2\sigma$ (the slope PDFs are essentially Gaussian distributions; Figures \ref{fig:PDF} and \ref{fig:PDF_KStest}). However, this comparison considers only the absolute value of the Salpeter slope without any associated uncertainty. From the Monte Carlo simulations shown in Figure \ref{fig:slope_flattening}, the uncertainty of the recovered slope from Salpeter distributions with sample sizes equal to that of the \textit{corrected} CMF in the mass range of our fits is $\approx$0.20. Considering this uncertainty on the Salpeter slope and the uncertainty of the derived G33.92 CMF, both slopes turn out to be consistent within $1\sigma$.

For a further, more robust comparison of the G33.92 CMF and the Salpeter IMF, we simulated populations following a Salpeter distribution ($\Gamma=-1.35$) and made them as incomplete as the \textit{raw} CMF (Figure \ref{fig:CMF_corrections}), as explained in Section \ref{sec:KS_test}. We carry out a two-sample KS test between the simulated incomplete samples and the \textit{raw} CMF. From the simulation of 10$^4$ samples we obtain a median $p$-value of 0.67, which can be interpreted as the probability of the null hypothesis being true, i.e., the probability that these distributions are drawn from the same parent sample. For such a high $p$-value the null hypothesis cannot be rejected and, therefore, we conclude that the high-mass end of the G33.92 CMF is statistically indistinguishable from the Salpeter stellar IMF.

\begin{table*}
\caption{CMFs of Diverse Star-Forming Regions Using ALMA Observations.}
  \scriptsize
  \centering
  \label{tab:cmf_comparisons}
  \begin{threeparttable}
	\begin{tabularx}{\linewidth}{@{\extracolsep{0pt}}lccccccc@{}}
    \toprule
	Region         & Mass Range   & $\Gamma$                & Synthesized Beam     & Distance & Resolution$^a$   & Detection Algorithm & Reference \\
	               & ($M_\odot$)  &                         & ($''\times''$)       & (kpc)    & (au)             &                     &           \\
	\midrule                                                                                    
    G33.92+0.11    & $>2.0$       & $-1.11_{-0.11}^{+0.12}$ & $0.16\times0.09$     & 7.1      & 1000             & \textit{dendrogram} & This work \\
    3 Clouds$^b$   & $>5.90$      & $-1.04\pm0.08$          & $0.25\times0.17$     & 8.2      & 1700             & \textit{dendrogram} & \citet{Lu_etal2020}         \\
	NGC6334        & $>2$         & $-1.10\pm0.02$          & $1.6\times1.2$       & 1.3      & 1800             & \texttt{SExtractor} & \citet{Sadaghiani_etal2020} \\
	W43-MM1        & 1.6--100     & $-0.96\pm0.02$          & $0.53\times0.37$     & 5.5      & 2400             & \textit{getsources} & \citet{Motte_etal2018}      \\
	G28.37+0.07    & $>0.79$      & $-0.87\pm0.07^c$        & $0.5\times0.5$       & 5        & 2500             & \textit{dendrogram} & \citet{Kong2019}            \\
	G28.37+0.07    & $>0.79$      & $-1.37\pm0.06^c$        & $0.5\times0.5$       & 5        & 2500             & \textit{astrograph} & \citet{Kong2019}            \\
	G286.21+0.17   & $\gtrsim1$   & $-1.24\pm0.17$          & $1.07\times1.02$     & 2.5      & 2600             & \textit{dendrogram} & \citet{Cheng_etal2018}      \\
	G286.21+0.17   & $\gtrsim1$   & $-0.64\pm0.13$          & $1.07\times1.02$     & 2.5      & 2600             & \textit{clumpfind}  & \citet{Cheng_etal2018}      \\
	28 dense clumps$^d$ & $\gtrsim5.0$ & $-0.94\pm0.08$     & $\approx 1\times1^e$ & 2.8$^e$  & $\approx 2800^f$ & \textit{dendrogram} & \citet{Oneill_etal2021}     \\
	7 IRDCs        & $\ge0.79$    & $-0.86\pm0.11$          & $1.36\times0.82$     & $4.4^e$  & $4600^f$         & \textit{dendrogram} & \citet{Liu_etal2018}        \\
	7 IRDCs        & $\ge1.26$    & $-0.70\pm0.13$          & $1.36\times0.82$     & $4.4^e$  & $4600^f$         & \textit{dendrogram} & \citet{Liu_etal2018}        \\
	12 IRDCs       & $\gtrsim0.6$ & $-1.07\pm0.09$          & $1.33\times1.13^e$   & $4^e$    & $4900^f$         & \textit{dendrogram} & \citet{Sanhueza_etal2019}   \\
	G305.137+0.069 & $>3$         & $0.02\pm0.37$           & $2.39\times2.10$     & 3.4      & 7600             & GaussClump          & \citet{Servajean_etal2019}  \\
    \bottomrule
	\end{tabularx}
	\begin{tablenotes}[para,flushleft]
	    $^a$ Considering the geometric average of the beam.\\
	    $^b$ In the Central Molecular Zone of the Galaxy.\\
 	    $^c$ From the scenario where the NH$_3$ gas temperature is applied to the cores.\\
	    $^d$ ALMAGAL project. \\
	    $^e$ Average value.\\
	    $^f$ At the average distance.\\
	\end{tablenotes}
 \end{threeparttable}
\end{table*}

\section{Discussion}
\label{sec:discussion}
Our result that the G33.92 CMF is statistically indistinguishable from the Salpeter stellar IMF is consistent with the idea that the high-mass end of the stellar IMF could directly originate from the CMF, as suggested by previous observations  \citep[e.g.;][]{Alves_etal2007,Cheng_etal2018,Massi_etal2019,Takemura_etal2021} and simulations  \citep[e.g.;][]{Padoan-Nordlund2002,Chabrier-Hennebelle2010,Vazquez-Semadeni_etal2019}. However, the question remains open as to why different star formation regions appear to have somewhat different CMF slopes (Table \ref{tab:cmf_comparisons}). One of the main possibilities is time evolution of the CMF due to the appearance of physical ingredients such as stellar feedback. The physical resolution of the measurements could also play a role (Louvet et al., submitted). To our knowledge, in this paper we report the highest-resolution measurement to date of a CMF.

If the flattening bias affecting the slope of the best power-law fits to the CMF were not corrected (Table \ref{tab:fitted_slopes}), one could be misled to conclude that the CMF slope differs from the stellar IMF by more than $4\sigma$. Thus, we stress the significance of taking into account these flattening effects on least-squares fittings of power-law functions to any kind of distributions that are represented by KDEs or histograms. For the case of differential distributions it is important mainly for small samples, while for continuous distributions represented by KDEs, the bias could be less pronounced (or might even disappear), but it affects small and large samples in a similar way (Figure \ref{fig:slope_flattening}).

We compared our high-mass CMF to the Salpeter IMF because of its validity for field stars in the solar neighborhood and in a variety of stellar associations \citep[e.g.;][]{Bastian_etal2010,Suarez_etal2019,Muzic_etal2019}. This has been interpreted as a suggestion of a universal IMF \citep{Bastian_etal2010,Offner_etal2014}. However, there is recent evidence of variations in the high-mass IMF. For example, \citet{Schneider_etal2018} reports a flatter slope  $\Gamma=-0.90_{-0.26}^{+0.37}$ in the extreme star forming environment of 30 Doradus in the Large Magellanic Cloud. Similarly, \citet{Zhang_etal2018} finds evidence for a top-heavy IMF in starburst galaxies from CO line ratios. In contrast, \citet{Weisz_etal2015} finds a steeper IMF slope $\Gamma=-1.45_{-0.06}^{+0.03}$ in a sample of young intermediate-mass star clusters. For a more detailed discussion on the possible (or not) universality of the IMF and its relation to the CMF, we refer the reader to the reviews by \citet{Bastian_etal2010}, \citet{Offner_etal2014}, \citet{Hopkins2018}, \citet{Kroupa2020}, and \citet{Lee_etal2020}.

\section{Summary}
\label{sec:summary}
We present the high-mass CMF of the G33.92 massive clump using high-resolution ($\approx 1000$ au) ALMA observations. This is one of the highest physical-resolution CMF measurements to date (Table \ref{tab:cmf_comparisons}). The CMF is based on 40 cores detected using the \textit{dendrogram} algorithm (Figure \ref{fig:structure_cores} and Table \ref{tab:cores_params}). We estimated the incompleteness of the sample (Figure \ref{fig:CMF_corrections}) by running Monte Carlo simulations to insert and recover artificial cores on the ALMA images. This allows us to convert the observed or \textit{raw} CMF into the \textit{corrected} CMF (Figure \ref{fig:CMF_final}), which is considered complete in the mass range $0.3\le \log(M/M_\odot)\le1.9$.

We followed two different approaches to find the best slope representation of the CMF: $i)$ fitting of power-law functions to the \textit{corrected} CMF using the least-squares method (Section \ref{sec:slope_fits}) and $ii)$ by comparing the \textit{raw} CMF to simulated incomplete samples using KS tests (Section \ref{sec:KS_test}). In the first case, which is the most widely used in the literature, the slopes of the best power-law fits suffer from the flattening biases shown in Figure \ref{fig:slope_flattening}, which are caused by the small size of the sample (when the CMF is built using classical histograms; Section \ref{sec:small-sample_correction}) or by edge effects (when using KDEs to represent the CMF; Section \ref{sec:edge_effects}). We run Monte Carlo simulations to correct such flattening effects on the slope fits. The inferred slopes (Figures \ref{fig:PDF} and \ref{fig:PDF_KStest}) from both approaches are in agreement with each other after correcting for the respective bias, as listed in Table \ref{tab:fitted_slopes} and shown in Figure \ref{fig:CMF_final}. The second approach, which does not consider slope fits, leads to a consistent and slightly more precise slope  $\Gamma=-1.11_{-0.12}^{+0.11}$, which we take as the slope for the CMF of G33.92.

The resulting CMF is consistent with those of other massive star forming regions, but not all (Table \ref{tab:cmf_comparisons}). Some of these regions have been reported to have an excess of high-mass cores with respect to the Salpeter stellar IMF. However, by carrying out KS tests between our observed CMF and simulated incomplete distributions from the Salpeter IMF ($\Gamma=-1.35$; Section \ref{sec:comparisons_CMF_IMF}), we conclude that the high-mass CMF of G33.92 is statistically indistinguishable from the (Salpeter) stellar IMF. This result is consistent with the idea that the shape of the IMF is inherited from the CMF to some important degree, but the question remains open if there is a time evolution of the CMF, or factors such as physical resolution have an effect on CMF measurements.

\bigskip

\acknowledgments 
The authors acknowledge support from UNAM-PAPIIT project IN104319. G.S., R.G.-M., and C.R.-Z. also acknowledge funding support from CONACyT Ciencia de Frontera project ID: 86372 (Citlacoatl). C.R.-Z. acknowledges support from project CONACyT CB2017-2018 AS-1-9754 Mexico. 

This paper makes use of the following ALMA data: ADS/JAO.ALMA\#2012.1.00387.S, ADS/JAO.ALMA\#2016.1.00362. ALMA is a partnership of ESO (representing its member states), NSF (USA) and NINS (Japan), together with NRC (Canada), MOST and ASIAA (Taiwan), and KASI (Republic of Korea), in cooperation with the Republic of Chile. The Joint ALMA Observatory is operated by ESO, AUI/NRAO and NAOJ. The National Radio Astronomy Observatory is a facility of the National Science Foundation operated undercooperative agreement by Associated Universities, Inc.

\bigskip

\facilities{ALMA}
\software{Astropy \citep{AstropyCollaboration2018}, SciPy \citep{SciPy-NMeth2020}, Astrodendro (http://www.dendrograms.org)}

\bibliography{mybib_Suarez}

\begin{thebibliography}{}
\expandafter\ifx\csname natexlab\endcsname\relax\def\natexlab#1{#1}\fi
\providecommand{\url}[1]{\href{#1}{#1}}
\providecommand{\dodoi}[1]{doi:~\href{http://doi.org/#1}{\nolinkurl{#1}}}
\providecommand{\doeprint}[1]{\href{http://ascl.net/#1}{\nolinkurl{http://ascl.net/#1}}}
\providecommand{\doarXiv}[1]{\href{https://arxiv.org/abs/#1}{\nolinkurl{https://arxiv.org/abs/#1}}}

\bibitem[{{Alves} {et~al.}(2007){Alves}, {Lombardi}, \&
  {Lada}}]{Alves_etal2007}
{Alves}, J., {Lombardi}, M., \& {Lada}, C.~J. 2007, \aap, 462, L17,
  \dodoi{10.1051/0004-6361:20066389}

\bibitem[{{Astropy Collaboration} {et~al.}(2018){Astropy Collaboration},
  {Price-Whelan}, {Sip{\H{o}}cz}, {G{\"u}nther}, {Lim}, {Crawford}, {Conseil},
  {Shupe}, {Craig}, {Dencheva}, {Ginsburg}, {Vand erPlas}, {Bradley},
  {P{\'e}rez-Su{\'a}rez}, {de Val-Borro}, {Aldcroft}, {Cruz}, {Robitaille},
  {Tollerud}, {Ardelean}, {Babej}, {Bach}, {Bachetti}, {Bakanov}, {Bamford},
  {Barentsen}, {Barmby}, {Baumbach}, {Berry}, {Biscani}, {Boquien}, {Bostroem},
  {Bouma}, {Brammer}, {Bray}, {Breytenbach}, {Buddelmeijer}, {Burke},
  {Calderone}, {Cano Rodr{\'\i}guez}, {Cara}, {Cardoso}, {Cheedella}, {Copin},
  {Corrales}, {Crichton}, {D'Avella}, {Deil}, {Depagne}, {Dietrich}, {Donath},
  {Droettboom}, {Earl}, {Erben}, {Fabbro}, {Ferreira}, {Finethy}, {Fox},
  {Garrison}, {Gibbons}, {Goldstein}, {Gommers}, {Greco}, {Greenfield},
  {Groener}, {Grollier}, {Hagen}, {Hirst}, {Homeier}, {Horton}, {Hosseinzadeh},
  {Hu}, {Hunkeler}, {Ivezi{\'c}}, {Jain}, {Jenness}, {Kanarek}, {Kendrew},
  {Kern}, {Kerzendorf}, {Khvalko}, {King}, {Kirkby}, {Kulkarni}, {Kumar},
  {Lee}, {Lenz}, {Littlefair}, {Ma}, {Macleod}, {Mastropietro}, {McCully},
  {Montagnac}, {Morris}, {Mueller}, {Mumford}, {Muna}, {Murphy}, {Nelson},
  {Nguyen}, {Ninan}, {N{\"o}the}, {Ogaz}, {Oh}, {Parejko}, {Parley}, {Pascual},
  {Patil}, {Patil}, {Plunkett}, {Prochaska}, {Rastogi}, {Reddy Janga},
  {Sabater}, {Sakurikar}, {Seifert}, {Sherbert}, {Sherwood-Taylor}, {Shih},
  {Sick}, {Silbiger}, {Singanamalla}, {Singer}, {Sladen}, {Sooley},
  {Sornarajah}, {Streicher}, {Teuben}, {Thomas}, {Tremblay}, {Turner},
  {Terr{\'o}n}, {van Kerkwijk}, {de la Vega}, {Watkins}, {Weaver}, {Whitmore},
  {Woillez}, {Zabalza}, \& {Astropy Contributors}}]{AstropyCollaboration2018}
{Astropy Collaboration}, {Price-Whelan}, A.~M., {Sip{\H{o}}cz}, B.~M., {et~al.}
  2018, \aj, 156, 123, \dodoi{10.3847/1538-3881/aabc4f}

\bibitem[{{Ballesteros-Paredes} {et~al.}(2015){Ballesteros-Paredes},
  {Hartmann}, {P{\'e}rez-Goytia}, \&
  {Kuznetsova}}]{BallesterosParedes_etal2015}
{Ballesteros-Paredes}, J., {Hartmann}, L.~W., {P{\'e}rez-Goytia}, N., \&
  {Kuznetsova}, A. 2015, \mnras, 452, 566, \dodoi{10.1093/mnras/stv1285}

\bibitem[{{Bastian} {et~al.}(2010){Bastian}, {Covey}, \&
  {Meyer}}]{Bastian_etal2010}
{Bastian}, N., {Covey}, K.~R., \& {Meyer}, M.~R. 2010, \araa, 48, 339,
  \dodoi{10.1146/annurev-astro-082708-101642}

\bibitem[{{Bergin} \& {Tafalla}(2007)}]{Bergin-Tafalla2007}
{Bergin}, E.~A., \& {Tafalla}, M. 2007, \araa, 45, 339,
  \dodoi{10.1146/annurev.astro.45.071206.100404}

\bibitem[{{Beuther} {et~al.}(2018){Beuther}, {Mottram}, {Ahmadi}, {Bosco},
  {Linz}, {Henning}, {Klaassen}, {Winters}, {Maud}, {Kuiper}, {Semenov},
  {Gieser}, {Peters}, {Urquhart}, {Pudritz}, {Ragan}, {Feng}, {Keto},
  {Leurini}, {Cesaroni}, {Beltran}, {Palau}, {S{\'a}nchez-Monge},
  {Galvan-Madrid}, {Zhang}, {Schilke}, {Wyrowski}, {Johnston}, {Longmore},
  {Lumsden}, {Hoare}, {Menten}, \& {Csengeri}}]{Beuther_etal2018}
{Beuther}, H., {Mottram}, J.~C., {Ahmadi}, A., {et~al.} 2018, \aap, 617, A100,
  \dodoi{10.1051/0004-6361/201833021}

\bibitem[{{Bonnell} {et~al.}(2011){Bonnell}, {Smith}, {Clark}, \&
  {Bate}}]{Bonnell_etal2011}
{Bonnell}, I.~A., {Smith}, R.~J., {Clark}, P.~C., \& {Bate}, M.~R. 2011,
  \mnras, 410, 2339, \dodoi{10.1111/j.1365-2966.2010.17603.x}

\bibitem[{{Chabrier} \& {Hennebelle}(2010)}]{Chabrier-Hennebelle2010}
{Chabrier}, G., \& {Hennebelle}, P. 2010, \apjl, 725, L79,
  \dodoi{10.1088/2041-8205/725/1/L79}

\bibitem[{{Cheng} {et~al.}(2018){Cheng}, {Tan}, {Liu}, {Kong}, {Lim},
  {Andersen}, \& {Da Rio}}]{Cheng_etal2018}
{Cheng}, Y., {Tan}, J.~C., {Liu}, M., {et~al.} 2018, \apj, 853, 160,
  \dodoi{10.3847/1538-4357/aaa3f1}

\bibitem[{{Clauset} {et~al.}(2009){Clauset}, {Shalizi}, \&
  {Newman}}]{Clauset_etal2009}
{Clauset}, A., {Shalizi}, C.~R., \& {Newman}, M.~E.~J. 2009, SIAM Review, 51,
  661, \dodoi{10.1137/070710111}

\bibitem[{{Draine}(2006)}]{Draine2006}
{Draine}, B.~T. 2006, \apj, 636, 1114, \dodoi{10.1086/498130}

\bibitem[{{Fish} {et~al.}(2003){Fish}, {Reid}, {Wilner}, \&
  {Churchwell}}]{Fish_etal2003}
{Fish}, V.~L., {Reid}, M.~J., {Wilner}, D.~J., \& {Churchwell}, E. 2003, \apj,
  587, 701, \dodoi{10.1086/368284}

\bibitem[{{Hennebelle} \& {Chabrier}(2008)}]{Hennebelle-Chabrier2008}
{Hennebelle}, P., \& {Chabrier}, G. 2008, \apj, 684, 395,
  \dodoi{10.1086/589916}

\bibitem[{{Hopkins}(2018)}]{Hopkins2018}
{Hopkins}, A.~M. 2018, \pasa, 35, \dodoi{10.1017/pasa.2018.29}

\bibitem[{{Izquierdo} {et~al.}(2018){Izquierdo}, {Galv{\'a}n-Madrid}, {Maud},
  {Hoare}, {Johnston}, {Keto}, {Zhang}, \& {de Wit}}]{Izquierdo_etal2018}
{Izquierdo}, A.~F., {Galv{\'a}n-Madrid}, R., {Maud}, L.~T., {et~al.} 2018,
  \mnras, 478, 2505, \dodoi{10.1093/mnras/sty1096}

\bibitem[{{Koen} \& {Kondlo}(2009)}]{Koen-Kondlo2009}
{Koen}, C., \& {Kondlo}, L. 2009, \mnras, 397, 495,
  \dodoi{10.1111/j.1365-2966.2009.14956.x}

\bibitem[{{Kong}(2019)}]{Kong2019}
{Kong}, S. 2019, \apj, 873, 31, \dodoi{10.3847/1538-4357/aaffd5}

\bibitem[{{K{\"o}nyves} {et~al.}(2015){K{\"o}nyves}, {Andr{\'e}},
  {Men'shchikov}, {Palmeirim}, {Arzoumanian}, {Schneider}, {Roy}, {Didelon},
  {Maury}, {Shimajiri}, {Di Francesco}, {Bontemps}, {Peretto}, {Benedettini},
  {Bernard}, {Elia}, {Griffin}, {Hill}, {Kirk}, {Ladjelate}, {Marsh}, {Martin},
  {Motte}, {Nguy{\^e}n Luong}, {Pezzuto}, {Roussel}, {Rygl}, {Sadavoy},
  {Schisano}, {Spinoglio}, {Ward-Thompson}, \& {White}}]{Konyves_etal2015}
{K{\"o}nyves}, V., {Andr{\'e}}, P., {Men'shchikov}, A., {et~al.} 2015, \aap,
  584, A91, \dodoi{10.1051/0004-6361/201525861}

\bibitem[{{Kroupa}(2020)}]{Kroupa2020}
{Kroupa}, P. 2020, in Star Clusters: From the Milky Way to the Early Universe,
  ed. A.~{Bragaglia}, M.~{Davies}, A.~{Sills}, \& E.~{Vesperini}, Vol. 351,
  117--121, \dodoi{10.1017/S1743921319007749}

\bibitem[{{Krumholz} {et~al.}(2012){Krumholz}, {Klein}, \&
  {McKee}}]{Krumholz_etal2012}
{Krumholz}, M.~R., {Klein}, R.~I., \& {McKee}, C.~F. 2012, \apj, 754, 71,
  \dodoi{10.1088/0004-637X/754/1/71}

\bibitem[{{Lee} {et~al.}(2020){Lee}, {Offner}, {Hennebelle}, {Andr{\'e}},
  {Zinnecker}, {Ballesteros-Paredes}, {Inutsuka}, \&
  {Kruijssen}}]{Lee_etal2020}
{Lee}, Y.-N., {Offner}, S. S.~R., {Hennebelle}, P., {et~al.} 2020, \ssr, 216,
  70, \dodoi{10.1007/s11214-020-00699-2}

\bibitem[{{Liu} {et~al.}(2015){Liu}, {Galv{\'a}n-Madrid}, {Jim{\'e}nez-Serra},
  {Rom{\'a}n-Z{\'u}{\~n}iga}, {Zhang}, {Li}, \& {Chen}}]{Liu_etal2015}
{Liu}, H.~B., {Galv{\'a}n-Madrid}, R., {Jim{\'e}nez-Serra}, I., {et~al.} 2015,
  \apj, 804, 37, \dodoi{10.1088/0004-637X/804/1/37}

\bibitem[{{Liu} {et~al.}(2012){Liu}, {Jim{\'e}nez-Serra}, {Ho}, {Chen},
  {Zhang}, \& {Li}}]{Liu_etal2012}
{Liu}, H.~B., {Jim{\'e}nez-Serra}, I., {Ho}, P. T.~P., {et~al.} 2012, \apj,
  756, 10, \dodoi{10.1088/0004-637X/756/1/10}

\bibitem[{{Liu} {et~al.}(2019){Liu}, {Chen}, {Rom{\'a}n-Z{\'u}{\~n}iga},
  {Galv{\'a}n-Madrid}, {Ginsburg}, {Ho}, {Minh}, {Jim{\'e}nez-Serra}, {Testi},
  \& {Zhang}}]{Liu_etal2019}
{Liu}, H.~B., {Chen}, H.-R.~V., {Rom{\'a}n-Z{\'u}{\~n}iga}, C.~G., {et~al.}
  2019, \apj, 871, 185, \dodoi{10.3847/1538-4357/aaf6b4}

\bibitem[{{Liu} {et~al.}(2018){Liu}, {Tan}, {Cheng}, \& {Kong}}]{Liu_etal2018}
{Liu}, M., {Tan}, J.~C., {Cheng}, Y., \& {Kong}, S. 2018, \apj, 862, 105,
  \dodoi{10.3847/1538-4357/aacb7c}

\bibitem[{{Lu} {et~al.}(2020){Lu}, {Cheng}, {Ginsburg}, {Longmore},
  {Kruijssen}, {Battersby}, {Zhang}, \& {Walker}}]{Lu_etal2020}
{Lu}, X., {Cheng}, Y., {Ginsburg}, A., {et~al.} 2020, \apjl, 894, L14,
  \dodoi{10.3847/2041-8213/ab8b65}

\bibitem[{{Ma{\'{\i}}z Apell{\'a}niz} \&
  {{\'U}beda}(2005)}]{MaizApellaniz-Ubeda2005}
{Ma{\'{\i}}z Apell{\'a}niz}, J., \& {{\'U}beda}, L. 2005, \apj, 629, 873,
  \dodoi{10.1086/431458}

\bibitem[{{Massi} {et~al.}(2019){Massi}, {Weiss}, {Elia}, {Csengeri},
  {Schisano}, {Giannini}, {Hill}, {Lorenzetti}, {Menten}, {Olmi}, {Schuller},
  {Strafella}, {De Luca}, {Motte}, \& {Wyrowski}}]{Massi_etal2019}
{Massi}, F., {Weiss}, A., {Elia}, D., {et~al.} 2019, \aap, 628, A110,
  \dodoi{10.1051/0004-6361/201935047}

\bibitem[{{Motte} {et~al.}(1998){Motte}, {Andre}, \& {Neri}}]{Motte_etal1998}
{Motte}, F., {Andre}, P., \& {Neri}, R. 1998, \aap, 336, 150

\bibitem[{{Motte} {et~al.}(2018{\natexlab{a}}){Motte}, {Bontemps}, \&
  {Louvet}}]{Motte_etal2018b}
{Motte}, F., {Bontemps}, S., \& {Louvet}, F. 2018{\natexlab{a}}, \araa, 56, 41,
  \dodoi{10.1146/annurev-astro-091916-055235}

\bibitem[{{Motte} {et~al.}(2018{\natexlab{b}}){Motte}, {Nony}, {Louvet},
  {Marsh}, {Bontemps}, {Whitworth}, {Men'shchikov}, {Nguyen Luong}, {Csengeri},
  {Maury}, {Gusdorf}, {Chapillon}, {K{\"o}nyves}, {Schilke}, {Duarte-Cabral},
  {Didelon}, \& {Gaudel}}]{Motte_etal2018}
{Motte}, F., {Nony}, T., {Louvet}, F., {et~al.} 2018{\natexlab{b}}, Nature
  Astronomy, 2, 478, \dodoi{10.1038/s41550-018-0452-x}

\bibitem[{{Mu{\v{z}}i{\'c}} {et~al.}(2019){Mu{\v{z}}i{\'c}}, {Scholz},
  {Pe{\~n}a Ram{\'\i}rez}, {Jayawardhana}, {Sch{\"o}del}, {Geers}, {Cieza}, \&
  {Bayo}}]{Muzic_etal2019}
{Mu{\v{z}}i{\'c}}, K., {Scholz}, A., {Pe{\~n}a Ram{\'\i}rez}, K., {et~al.}
  2019, \apj, 881, 79, \dodoi{10.3847/1538-4357/ab2da4}

\bibitem[{{Nony} {et~al.}(2020){Nony}, {Motte}, {Louvet}, {Plunkett},
  {Gusdorf}, {Fechtenbaum}, {Pouteau}, {Lefloch}, {Bontemps}, {Molet}, \&
  {Robitaille}}]{Nony_etal2020}
{Nony}, T., {Motte}, F., {Louvet}, F., {et~al.} 2020, \aap, 636, A38,
  \dodoi{10.1051/0004-6361/201937046}

\bibitem[{{Offner} {et~al.}(2014){Offner}, {Clark}, {Hennebelle}, {Bastian},
  {Bate}, {Hopkins}, {Moraux}, \& {Whitworth}}]{Offner_etal2014}
{Offner}, S.~S.~R., {Clark}, P.~C., {Hennebelle}, P., {et~al.} 2014, Protostars
  and Planets VI, 53, \dodoi{10.2458/azu_uapress_9780816531240-ch003}

\bibitem[{{O'Neill} {et~al.}(2021){O'Neill}, {Cosentino}, {Tan}, {Cheng}, \&
  {Liu}}]{Oneill_etal2021}
{O'Neill}, T.~J., {Cosentino}, G., {Tan}, J.~C., {Cheng}, Y., \& {Liu}, M.
  2021, arXiv e-prints, arXiv:2104.08861.
\newblock \doarXiv{2104.08861}

\bibitem[{{Ossenkopf} \& {Henning}(1994)}]{Ossenkopf-Henning_1994}
{Ossenkopf}, V., \& {Henning}, T. 1994, \aap, 291, 943

\bibitem[{{Padoan} \& {Nordlund}(2002)}]{Padoan-Nordlund2002}
{Padoan}, P., \& {Nordlund}, {\AA}. 2002, \apj, 576, 870,
  \dodoi{10.1086/341790}

\bibitem[{{Palau} {et~al.}(2018){Palau}, {Zapata}, {Rom{\'a}n-Z{\'u}{\~n}iga},
  {S{\'a}nchez-Monge}, {Estalella}, {Busquet}, {Girart}, {Fuente}, \&
  {Commer{\c{c}}on}}]{Palau_etal2018}
{Palau}, A., {Zapata}, L.~A., {Rom{\'a}n-Z{\'u}{\~n}iga}, C.~G., {et~al.} 2018,
  \apj, 855, 24, \dodoi{10.3847/1538-4357/aaad03}

\bibitem[{{Pineda} {et~al.}(2009){Pineda}, {Rosolowsky}, \&
  {Goodman}}]{Pineda_etal2009}
{Pineda}, J.~E., {Rosolowsky}, E.~W., \& {Goodman}, A.~A. 2009, \apjl, 699,
  L134, \dodoi{10.1088/0004-637X/699/2/L134}

\bibitem[{{Press} {et~al.}(1992){Press}, {Teukolsky}, {Vetterling}, \&
  {Flannery}}]{Press_etal1992}
{Press}, W.~H., {Teukolsky}, S.~A., {Vetterling}, W.~T., \& {Flannery}, B.~P.
  1992, {Numerical recipes in FORTRAN. The art of scientific computing}

\bibitem[{{Rosolowsky} {et~al.}(2008){Rosolowsky}, {Pineda}, {Kauffmann}, \&
  {Goodman}}]{Rosolowsky_etal2008}
{Rosolowsky}, E.~W., {Pineda}, J.~E., {Kauffmann}, J., \& {Goodman}, A.~A.
  2008, \apj, 679, 1338, \dodoi{10.1086/587685}

\bibitem[{{Sadaghiani} {et~al.}(2020){Sadaghiani}, {S{\'a}nchez-Monge},
  {Schilke}, {Liu}, {Clarke}, {Zhang}, {Girart}, {Seifried}, {Aghababaei},
  {Li}, {Ju{\'a}rez}, \& {Tang}}]{Sadaghiani_etal2020}
{Sadaghiani}, M., {S{\'a}nchez-Monge}, {\'A}., {Schilke}, P., {et~al.} 2020,
  \aap, 635, A2, \dodoi{10.1051/0004-6361/201935699}

\bibitem[{{Salpeter}(1955)}]{Salpeter1955}
{Salpeter}, E.~E. 1955, \apj, 121, 161, \dodoi{10.1086/145971}

\bibitem[{{Sanhueza} {et~al.}(2019){Sanhueza}, {Contreras}, {Wu}, {Jackson},
  {Guzm{\'a}n}, {Zhang}, {Li}, {Lu}, {Silva}, {Izumi}, {Liu}, {Miura},
  {Tatematsu}, {Sakai}, {Beuther}, {Garay}, {Ohashi}, {Saito}, {Nakamura},
  {Saigo}, {Veena}, {Nguyen-Luong}, \& {Tafoya}}]{Sanhueza_etal2019}
{Sanhueza}, P., {Contreras}, Y., {Wu}, B., {et~al.} 2019, \apj, 886, 102,
  \dodoi{10.3847/1538-4357/ab45e9}

\bibitem[{{Schneider} {et~al.}(2018){Schneider}, {Sana}, {Evans},
  {Bestenlehner}, {Castro}, {Fossati}, {Gr{\"a}fener}, {Langer},
  {Ram{\'\i}rez-Agudelo}, {Sab{\'\i}n-Sanjuli{\'a}n}, {Sim{\'o}n-D{\'\i}az},
  {Tramper}, {Crowther}, {de Koter}, {de Mink}, {Dufton}, {Garcia}, {Gieles},
  {H{\'e}nault-Brunet}, {Herrero}, {Izzard}, {Kalari}, {Lennon}, {Ma{\'\i}z
  Apell{\'a}niz}, {Markova}, {Najarro}, {Podsiadlowski}, {Puls}, {Taylor}, {van
  Loon}, {Vink}, \& {Norman}}]{Schneider_etal2018}
{Schneider}, F.~R.~N., {Sana}, H., {Evans}, C.~J., {et~al.} 2018, Science, 359,
  69, \dodoi{10.1126/science.aan0106}

\bibitem[{{Servajean} {et~al.}(2019){Servajean}, {Garay}, {Rathborne},
  {Contreras}, \& {Gomez}}]{Servajean_etal2019}
{Servajean}, E., {Garay}, G., {Rathborne}, J., {Contreras}, Y., \& {Gomez}, L.
  2019, \apj, 878, 146, \dodoi{10.3847/1538-4357/ab204c}

\bibitem[{{Silverman}(1986)}]{Silverman1986}
{Silverman}, B.~W. 1986, {Density estimation for statistics and data analysis}

\bibitem[{{Su{\'a}rez} {et~al.}(2019){Su{\'a}rez}, {Downes},
  {Rom{\'a}n-Z{\'u}{\~n}iga}, {Cervi{\~n}o}, {Brice{\~n}o}, {Petr-Gotzens}, \&
  {Vivas}}]{Suarez_etal2019}
{Su{\'a}rez}, G., {Downes}, J.~J., {Rom{\'a}n-Z{\'u}{\~n}iga}, C., {et~al.}
  2019, \mnras, 486, 1718, \dodoi{10.1093/mnras/stz756}

\bibitem[{{Takemura} {et~al.}(2021){Takemura}, {Nakamura}, {Kong}, {Arce},
  {Carpenter}, {Ossenkopf-Okada}, {Klessen}, {Sanhueza}, {Shimajiri},
  {Tsukagoshi}, {Kawabe}, {Ishii}, {Dobashi}, {Shimoikura}, {Goldsmith},
  {S{\'a}nchez-Monge}, {Kauffmann}, {Pillai}, {Padoan}, {Ginsberg}, {Smith},
  {Bally}, {Mairs}, {Pineda}, {Lis}, {Burkhart}, {Schilke}, {Chen}, {Isella},
  {Friesen}, {Goodman}, \& {Harper}}]{Takemura_etal2021}
{Takemura}, H., {Nakamura}, F., {Kong}, S., {et~al.} 2021, arXiv e-prints,
  arXiv:2103.08527.
\newblock \doarXiv{2103.08527}

\bibitem[{{V{\'a}zquez-Semadeni} {et~al.}(2019){V{\'a}zquez-Semadeni}, {Palau},
  {Ballesteros-Paredes}, {G{\'o}mez}, \&
  {Zamora-Avil{\'e}s}}]{Vazquez-Semadeni_etal2019}
{V{\'a}zquez-Semadeni}, E., {Palau}, A., {Ballesteros-Paredes}, J.,
  {G{\'o}mez}, G.~C., \& {Zamora-Avil{\'e}s}, M. 2019, \mnras, 490, 3061,
  \dodoi{10.1093/mnras/stz2736}

\bibitem[{Virtanen {et~al.}(2020)Virtanen, Gommers, Oliphant, Haberland, Reddy,
  Cournapeau, Burovski, Peterson, Weckesser, Bright, {van der Walt}, Brett,
  Wilson, Millman, Mayorov, Nelson, Jones, Kern, Larson, Carey, Polat, Feng,
  Moore, {VanderPlas}, Laxalde, Perktold, Cimrman, Henriksen, Quintero, Harris,
  Archibald, Ribeiro, Pedregosa, {van Mulbregt}, \& {SciPy 1.0
  Contributors}}]{SciPy-NMeth2020}
Virtanen, P., Gommers, R., Oliphant, T.~E., {et~al.} 2020, Nature Methods, 17,
  261, \dodoi{10.1038/s41592-019-0686-2}

\bibitem[{{Wang} {et~al.}(2020){Wang}, {Koch}, {Galv{\'a}n-Madrid}, {Lai},
  {Liu}, {Lin}, \& {Pattle}}]{Wang_etal2020}
{Wang}, J.-W., {Koch}, P.~M., {Galv{\'a}n-Madrid}, R., {et~al.} 2020, arXiv
  e-prints, arXiv:2011.01555.
\newblock \doarXiv{2011.01555}

\bibitem[{{Watt} \& {Mundy}(1999)}]{Watt-Mundy1999}
{Watt}, S., \& {Mundy}, L.~G. 1999, \apjs, 125, 143, \dodoi{10.1086/313273}

\bibitem[{{Weisz} {et~al.}(2015){Weisz}, {Johnson}, {Foreman-Mackey},
  {Dolphin}, {Beerman}, {Williams}, {Dalcanton}, {Rix}, {Hogg}, {Fouesneau},
  {Johnson}, {Bell}, {Boyer}, {Gouliermis}, {Guhathakurta}, {Kalirai}, {Lewis},
  {Seth}, \& {Skillman}}]{Weisz_etal2015}
{Weisz}, D.~R., {Johnson}, L.~C., {Foreman-Mackey}, D., {et~al.} 2015, \apj,
  806, 198, \dodoi{10.1088/0004-637X/806/2/198}

\bibitem[{{Zhang} {et~al.}(2018){Zhang}, {Romano}, {Ivison}, {Papadopoulos}, \&
  {Matteucci}}]{Zhang_etal2018}
{Zhang}, Z.-Y., {Romano}, D., {Ivison}, R.~J., {Papadopoulos}, P.~P., \&
  {Matteucci}, F. 2018, \nat, 558, 260, \dodoi{10.1038/s41586-018-0196-x}

\end{thebibliography}
\bibliographystyle{aasjournal}

\end{document}